\documentclass[twocolumn]{aastex631}

\usepackage{amsmath}

\newcommand{\ergs}{~{\rm erg}~{\rm s}^{-1}}

\newcommand{\km}{\ {\rm km}}

\newcommand{\yr}{\ {\rm yr}}

\newcommand{\MeVnuc}{~{\rm MeV}~{\rm nuc^{-1}}}
\newcommand{\MeVnucs}{~{\rm MeV}~{\rm nuc^{-1}}~{\rm s^{-1}}}

\newcommand{\Ye}{Y_{\rm e}}

\newcommand{\wiadd}[1]{{#1}}
\shorttitle{Fallback Halted by $R$-process Heating}
\shortauthors{Ishizaki et al.}
\graphicspath{{./}{./}}

\begin{document}

\title{
	Fallback Accretion Halted by $R$-process Heating in Neutron Star Mergers and Gamma-Ray Bursts
}

\correspondingauthor{Wataru Ishizaki}
\email{wataru.ishizaki@yukawa.kyoto-u.ac.jp}

\author[0000-0002-7005-7139]{Wataru Ishizaki}
\affiliation{Center for Gravitational Physics, Yukawa Institute for Theoretical Physics, Kyoto University, Kyoto, 606-8502, Japan}

\author[0000-0003-4988-1438]{Kenta Kiuchi}
\affiliation{Center for Gravitational Physics, Yukawa Institute for Theoretical Physics, Kyoto University, Kyoto, 606-8502, Japan}
\affiliation{Max Planck Institute for Gravitational Physics (Albert Einstein Institute), Am M\"{u}hlenberg, Potsdam-Golm, 14476, Germany}

\author[0000-0002-3517-1956]{Kunihito Ioka}
\affiliation{Center for Gravitational Physics, Yukawa Institute for Theoretical Physics, Kyoto University, Kyoto, 606-8502, Japan}

\author[0000-0002-4759-7794]{Shinya Wanajo}
\affiliation{Max Planck Institute for Gravitational Physics (Albert Einstein Institute), Am M\"{u}hlenberg, Potsdam-Golm, 14476, Germany}
\affiliation{Interdisciplinary Theoretical and Mathematical Sciences Program (iTHEMS), RIKEN,
Wako, Saitama 351-0198, Japan}



\begin{abstract}
The gravitational wave event GW170817 with a macronova/kilonova shows that a merger of two neutron stars ejects matter with radioactivity including $r$-process nucleosynthesis.
A part of the ejecta inevitably falls back to the central object, possibly powering long-lasting activities of a short gamma-ray burst (sGRB), such as extended and plateau emissions.
We investigate the fallback accretion with the $r$-process heating by performing one-dimensional hydrodynamic simulations and developing a semi-analytical model.
We show that the usual fallback rate $dM/dt \propto t^{-5/3}$ is halted by the heating because pressure gradients accelerate ejecta beyond an escape velocity.
The suppression is steeper than Chevalier's power-law model through Bondi accretion within a turn-around radius.
The characteristic halting timescale is $\sim 10^4$--$10^8$ sec for the GW170817-like $r$-process heating, which is longer than the typical timescale of the long-lasting emission of sGRBs.
The halting timescale is sensitive to the uncertainty of the $r$-process.
Future observation of fallback halting could constrain the $r$-process heating on the year scale.
\end{abstract}



\section{INTRODUCTION}\label{sec:intro}

The discovery of the short-duration gamma-ray burst (sGRB), GRB 170817A, coincided with the detection of GW170817 by the Laser Interferometer Gravitational-Wave Observatory (LIGO) and the Virgo Consortium (LVC), is a direct evidence that binary neutron star (BNS) mergers are one of the sources of sGRBs \citep{2017ApJ...848L..13A,2017PhRvL.119p1101A}.
The simultaneous detection of macronova/kilonova emission is a strong indication for the ejection of neutron-rich matter and $r$-process element synthesis in this BNS merger \citep{2017Natur.551...64A,2017ApJ...848L..19C,2017Sci...358.1556C,2017ApJ...848L..17C,2017Sci...358.1570D,2017Natur.551...80K,2017Sci...358.1559K,2017Sci...358.1583K,2017ApJ...848L..32M,2017ApJ...848L..18N,2017Sci...358.1574S,2017Natur.551...75S,2017ApJ...848L..16S,2017PASJ...69..102T,2017ApJ...848L..27T}.
The observed properties of the gravitational waves and electromagnetic counterparts are broadly consistent with the predictions from a series of theoretical studies including numerical relativity calculations \citep{2017PhRvD..96l3012S,2017ApJ...848L..17C,2017Natur.551...80K,2017Sci...358.1559K,2017ApJ...851L..21V,2018PTEP.2018d3E02I,2019MNRAS.487.4884I} and these simultaneous detection has established the paradigm that sGRBs originate from compact binary mergers \citep{1986ApJ...308L..43P,1986ApJ...308L..47G,1989Natur.340..126E}.

Although sGRBs are classified as GRBs with the duration of prompt emission being less than $2\sec$ \citep{1993ApJ...413L.101K}, their central engines are thought to remain active for longer time \citep{2005Sci...309.1833B,2005ApJ...631..429I}.
This was suggested from the observations of extended emission which lasts for about $100 \sec$ with a luminosity of $10^{48}$--$10^{49}\ergs$ and possibly from the observations of plateau emission which lasts for about $10^4 \sec$ with a luminosity of $10^{45}$--$10^{46}\ergs$ \citep{2005SSRv..120..143B,2013MNRAS.430.1061R,2017ApJ...846..142K,2019ApJ...877..147K}.
These long-lasting emissions are fainter than the prompt emission in luminosity.
However, because of the long duration of these emissions, they are comparable with or greater than the prompt emission in fluence \citep{2017ApJ...846..142K}.
Some of the extended emissions have been observed to darken abruptly, which are not expected in the sGRB afterglow \citep{2005ApJ...631..429I}.
Therefore, a central engine activity is very likely to explain these long-lasting emission\footnote{Note that, although such a rapid fade-out has not been reported for the plateau emission, it may also reflect the late-time activity of the central engine \citep[see also][]{2020MNRAS.493..783M}.}.
There are two models for the energy source of the long-lasting emissions, which have been extensively discussed in previous works.
One is the release of rotational or magnetic field energy from a strongly magnetized massive neutron star that forms after a BNS merger \citep{2001ApJ...552L..35Z,2008MNRAS.385.1455M,2012MNRAS.419.1537B,2013ApJ...779L..25F,2018ApJ...854...60M}.
Currently, this scenario is not conclusive because the expected observational features, e.g., late radio emission, have not been detected \citep{2014MNRAS.437.1821M,2016ApJ...819L..22H}.

The other scenario for the energy source of long-lasting activity is the fallback accretion of ejecta \citep{2007MNRAS.376L..48R,2007NJPh....9...17L,2009MNRAS.392.1451R,2015ApJ...804L..16K,2017ApJ...846..142K}.
Numerical calculations of compact binary mergers show that a part of the ejecta is still gravitationally bound \citep[e.g.,][]{1999A&A...341..499R,2013ApJ...773...78B,2015PhRvD..92d4028K,2016MNRAS.460.3255R,2017PhRvD..96h4060K}.
\citet{2015PhRvD..92d4028K} calculated the coalescence of a black hole and a neutron star (BH--NS) based on numerical relativity and showed that a part of the ejecta falls back to the merger remnant.
While the gravitational energy released by fallback accretion is large enough to explain the extended emission and the plateau emission, the simple theory of fallback accretion, which assumes a zero-temperature fluid, predicts that the mass accretion rate is proportional to the power of time as $t^{-5/3}$ with no typical timescale \citep{1988Natur.333..523R,1988Natur.333..644M}.
However, the observed light curve of the long-lasting emission of sGRB clearly has a certain timescale, which is not compatible with the simple theory \citep{2019ApJ...877..147K}.

The coincidence of the macronova/kilonova emission with the gravitational wave source GW170817 indicates that the ejecta of the BNS merger is heated by the radioactive decay of $r$-process elements.
Therefore, the assumption of zero-temperature fluid is inappropriate \citep{2017Natur.551...75S,2018MNRAS.481.3423W,2018ApJ...865L..21K}.
\citet{2010MNRAS.406.2650M} discussed the effect of ejecta heating by the radioactive decay of $r$-process elements on the mass accretion using a test-particle model.
\citet{2019MNRAS.485.4404D} performed a more sophisticated calculation based on the model in \citet{2010MNRAS.406.2650M}, using the ejecta profiles obtained from numerical relativity simulations and the radioactive heating rates obtained from nucleosynthesis calculations.
They showed that the mass accretion stops and resumes after a finite time, the so-called ``gap'', because the marginally bound fluid elements become unbound by the heating due to the radioactive decay of $r$-process elements.
Furthermore, they argued that the timescale of this resumption is $\mathcal{O}(100)\sec$, which is in agreement with the timescale of the extended emission.

In the test-particle model of \citet{2010MNRAS.406.2650M} and \citet{2019MNRAS.485.4404D}, they assumed that all the radioactive energy is converted to the kinetic energy.
In reality, the energy from the radioactive heating is converted to the kinetic energy through the pressure gradient force resulting from the increased internal energy. 
It is unclear whether the pressure gradient is large enough to convert all the internal energy of the fluid element into the kinetic energy, so that the validity of their assumption is also unclear.
Because their assumption cannot be verified within the framework of the test-particle model, it is necessary to solve the hydrodynamic equations incorporating the effects of radioactive heating.

In this paper, we reconsider the effect of the radioactive heating due to decaying $r$-process elements on the fallback accretion in BNS mergers by numerically solving one-dimensional fluid equations.
We also construct a semi-analytical model that reproduces the hydrodynamical simulation results and explores it over a large parameter space including the realistic $r$-process heating.
This paper is organized as follows. 
In Section \ref{sec:CFD}, we describe the method of the numerical calculation of the fluid equations and show the results.
In Section \ref{sec:analytic} and Section \ref{sec:analytic_bPL}, the semi-analytical models of the accretion flow are developed for a constant heating rate and a heating rate of a broken power-law form, respectively.
In Section \ref{sec:application}, the semi-analytical model is applied to explore the parameters relevant to the radioactive heating.
In Section \ref{sec:summary}, we summarize this work, and discuss the scope of application of the spherically symmetric modeling used in this study and future prospects.

\section{Hydrodynamical Simulation of fallback accretion}\label{sec:CFD}

\subsection{Numerical method}
In order to investigate the effect of the radioactive heating due to $r$-process nuclei on the fallback accretion in the BNS merger,
we have performed long-term one-dimensional hydrodynamic simulations of the matter ejected during the merger.
The ejecta profiles of the velocity and the mass density are derived from the numerical relativity simulations performed by \citet{2017PhRvD..96h4060K}.
The hydrodynamical equations for spherically symmetric and purely radial flow, including the heating and the point source gravity, are written as follows:
\begin{equation}
\label{eq:eoc}
\frac{\partial \rho}{\partial t}+\frac{1}{r^{2}}\frac{\partial}{\partial r}\left(r^{2} \rho v\right)=0,
\end{equation}
\begin{equation}
\label{eq:momentum}
\frac{\partial v}{\partial t}+v \frac{\partial v}{\partial r}=-\frac{1}{\rho} \frac{\partial P}{\partial r}-\frac{G M}{r^{2}},
\end{equation}
\begin{multline}
\label{eq:energy}
\frac{\partial}{\partial t}\left(\frac{1}{2} \rho v^{2}+\epsilon_{\rm int}\right)
+\frac{1}{r^2}\frac{\partial }{\partial r}\left[r^2v\left(\frac{1}{2} \rho v^{2}+\epsilon_{\rm int}+P\right)\right] \\
=-\rho v \frac{GM}{r^2}+Q_{\rm heat},
\end{multline}
where 
$\rho$ is the mass density,
$v$ is the radial bulk velocity of the fluid,
$P$ is the pressure,
$\epsilon_{\rm int}$ is the internal energy,
$M$ is the mass of the central object,
and $Q_{\rm heat}$ is the radioactive heating rate per unit volume per unit time.\footnote{The radioactive heating due to the nuclear decay is the conversion of the mass deficit of the nucleus into the internal energy.
Therefore, the sum of the rest mass energy and the internal energy should be conserved, and it is not strictly correct to simply add the external heating term to the right-hand side of Equation (\ref{eq:energy}).
Here, we ignore the term $-Q_{\rm heat}/c^2$, which should be in the right-hand side of Equation (\ref{eq:eoc}).
In the decay of a nucleus, the ratio of the mass deficit to the nucleon mass is about $\mathcal{O}({\rm MeV})/\mathcal{O}({\rm GeV})$.
Since the heating due to this mass deficit is comparable to the internal energy, the magnitude of the term $-Q_{\rm heat}/c^2$ relative to the left hand side of Equation (\ref{eq:eoc}) is of the magnitude $\mathcal{O}(v^2/c^2)$.
In this paper, we ignore the $\mathcal{O}(v^2/c^2)$ term because we discuss the motion of ejecta in the Newtonian limit.
For a more formal formulation, see \citet{2017PhRvD..96h3016U}.}
The equation of state is assumed to follow the $\Gamma$-law,
\begin{equation}\label{eq:eos}
P=\left(\Gamma-1\right)\epsilon_{\rm int},
\end{equation}
where $\Gamma$ is the adiabatic index.
In Equation (\ref{eq:eos}), we adopt the radiation dominant case of $\Gamma=4/3$.
Here we neglect the radiative loss.
Furthermore, we also neglect the self-gravity of the ejecta, because the ejecta mass is much smaller than the central mass.

We solve Equations (\ref{eq:eoc})--(\ref{eq:eos}) by using a one-dimensional hydrodynamics code with the Newtonian gravity.
The advection term of the hydrodynamic equations is solved by using the HLL method \citep[e.g.,][]{2002A&A...390.1177D} and the 3rd-order MUSCL reconstruction \citep[see][for a review]{2017LRCA....3....2B}.
In the range of the parameters used in this paper, the system can be well described in a non-relativistic regime.
The calculation is performed by dividing the radius from $40 \km$ to $80,000\km$ into a uniform grid of 16,384 cells (about $\sim 4.9\km$ per grid).
The boundary conditions imposed on the inner and outer boundaries are such that the radial differential coefficient is set to zero for all physical quantities.
Note that, because the inner boundary is far inside from the sonic radius, the inner boundary condition corresponds to a free-flow condition to the central object.
In order to check the convergence, we perform the calculation with a spatial resolution that is twice as fine; as a result, the maximum relative error in the mass accretion rate is found to be about 0.2\% (the result is shown in Figure \ref{fig:mdot}).

\subsection{Initial profile of merger ejecta}\label{sec:initial}

\begin{figure}
	\centering
	\includegraphics[width=1.0\linewidth]{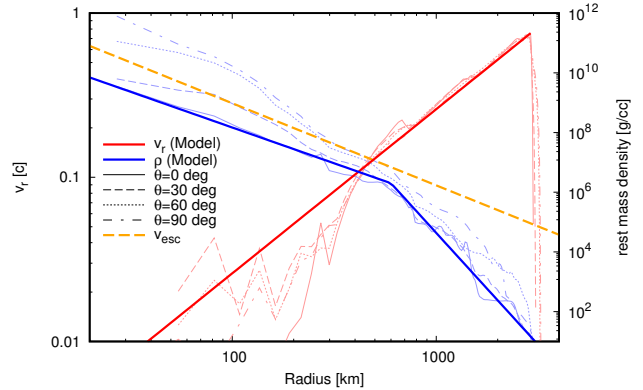}
	\caption{
		Radial profiles of the radial velocity (left axis) and the mass density (right axis).
		The thick lines represent the model curves at $t = 0$ used in the calculations (see text).
		The thin lines with different styles show the result of \citet{2017PhRvD..96h4060K} at $\approx 10$ ms after the merger for the zenith angles given in the legend.
		Note that only a profile of the dynamical ejecta is shown without a subsequent component (such as viscosity-driven wind).
		The bold yellow dashed line indicates the escape velocity.
		The region inside $\sim 490\km$ is gravitationally bound. 
	}
	\label{fig:ejecta}
\end{figure}

We model the initial profile of dynamical ejecta based on the results of the numerical relativity simulation of the BNS merger performed in \citet{2017PhRvD..96h4060K}. The mass for each NS is 1.35 $M_\odot$. The EOS of the NS matter is described by the two segments piece-wise polytropic EOS, which is referred to as the H model in \citet{2017PhRvD..96h4060K}. In this particular model, the radius of the NS with $1.35M_\odot$ is 12.3 km. 

Figure \ref{fig:ejecta} shows the radial profiles of the radial 3-velocity and the mass density.
The thin lines in Figure \ref{fig:ejecta} show the result of \citet{2017PhRvD..96h4060K} at $\approx 10$ ms after the merger for each zenith angle given in the legend.
The dashed yellow line in the figure shows the escape velocity from the gravitational field of the central object with mass $M=2.7M_\odot$. 
It can be confirmed that the fluid elements located at $r\lesssim 490\km$ are gravitationally bound.
Around the radius $r\sim490 \km$, which is important for the calculation of mass accretion rates, the radial dependence of velocity and density is found to be weekly dependent on latitude.
This is one justification of our treatment of spherically symmetric modelling.

In this paper, the initial velocity and density profiles shown in Figure \ref{fig:ejecta} are modeled as described below.
Given that the expansion of the dynamical ejecta can be regarded as a homologous expansion, we adopt a model in which the radial 3-velocity is proportional to the radius (out to $3000 \km$):
\begin{eqnarray}
v(t=0,r)=
\left\{
\begin{array}{ll}
\displaystyle 0.26\, c\left(\frac{r}{1000\km}\right) & \left(r<3000\km\right)\\
\displaystyle 0 & \left(r>3000\km\right)
\end{array}
\right.
,
\label{eq:vmodel}
\end{eqnarray}
where $t=0$ is set at the beginning of the fluid calculations in this study and corresponds to the time slice of the numerical relativity simulation in \citet{2017PhRvD..96h4060K} ($\approx 10$ ms after the merger).
The radius of $r\sim 3000\km$ corresponds to that the outermost ejecta with nearly the speed of light reach during about $10 ~{\rm ms}$.
For the mass density, we adopt a broken power-law model, which has a break at $600 \km$, namely
\begin{equation}
\rho(t=0,r)=
\left\{
\begin{array}{ll}
\displaystyle \rho_0 \left(\frac{r}{600\km}\right)^{-2.4} & \left(r<600\km\right)\vspace{1mm}\\
\displaystyle \rho_0 \left(\frac{r}{600\km}\right)^{-7.5} & \left(r>600\km\right)
\end{array}
\right.
,
\label{eq:rhomodel}
\end{equation} 
where $\rho_0=2.0\times 10^6~{\rm g~cm^{-3}}$ is the value at $r=600\km$. 
We assume $P=10^{-5}\rho c^2$ as the initial pressure distribution in the ejecta, because the initial internal energy is sufficiently low and does not affect the motion of the ejecta.

\citet{2020ApJ...901..122F} have pointed out that the viscosity-driven wind can be launched with the timescale of $\mathcal{O}(1) \sec$, which dominates the total mass of the ejecta from a BNS merger \citep[see also][]{2013MNRAS.435..502F,2015MNRAS.448..541J,2019MNRAS.482.3373F}.
\citet{2020arXiv201214711K} have studied further the long term temporal evolution of the viscosity-driven wind based on the results of \citet{2020ApJ...901..122F}.
According to these studies, after a time $\sim 10 \sec$, the velocity profile of the wind approximately matches that of a homologous expansion in Equation (\ref{eq:vmodel}). 
The density distribution is also expected to be close to $\rho\propto r^{-2}$ as that of a steady-state supersonic flow, which is qualitatively similar to the radial dependence of the density distribution expressed in Equation (\ref{eq:rhomodel}) for $r < 600$ km.
However, for the viscously driven wind, to take into account the difference in total ejecta mass, it will be necessary to increase the value of mass density $\rho_0$ by one order of magnitude compared with the value of the dynamical ejecta.
Here we presume that the disk wind is modeled by Equations (\ref{eq:vmodel}) and (\ref{eq:rhomodel}) with a larger value of density $\rho_0$ than that for the dynamical ejecta case.
Note that, at the time of interest for considering the effect of radioactive heating, the mass accretion rate is determined by the profile of the marginally bound ejecta rather than that of the overall ejecta.
Therefore, the detailed modeling of the mass density profile is probably not very important.

\subsection{Radioactive heating}

\begin{figure}
	\centering
	\includegraphics[width=1.0\linewidth]{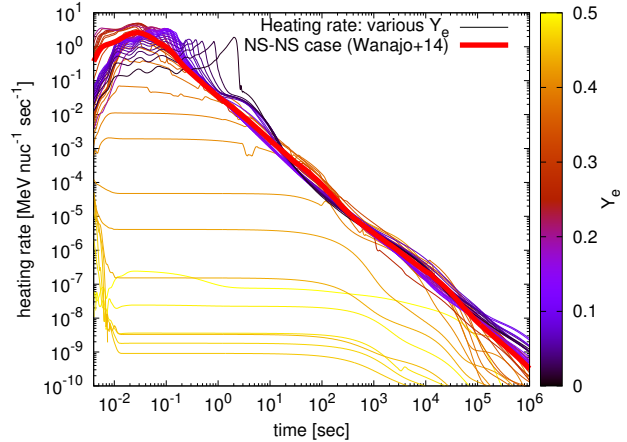}
	\caption{Temporal evolution of the radioactive heating rates (excluding the energies in neutrinos) adopted from \citet{2014ApJ...789L..39W}.
		The thick line represents the mass-averaged heating rate.
		For the fluid elements of various electron fractions $\Ye$, the respective heating rate is plotted as a thin line with the color indicated by the color bar. 
	}
	\label{fig:rprocess}
\end{figure}

Figure \ref{fig:rprocess} plots the radioactive heating rates adopted from the results of nucleosynthesis calculations based on the numerical model of a BNS merger \citep{2014ApJ...789L..39W}. The heating is due to $\beta$-decay, $\alpha$-decay, and fission of $r$-process nuclei produced in the dynamical ejecta. Each thin curve shows the heating rate in units of MeV nuc$^{-1}$ s$^{-1}$ as a function of time (since the merger) for the Lagrangian tracer-particle of the ejecta with a given initial $Y_\mathrm{e}$. The color indicates the value of $Y_\mathrm{e}$ from 0.09 (purple) to 0.44 (yellow) with an interval of $\Delta Y_\mathrm{e} = 0.01$. The heating rate averaged over the ejecta mass is also shown by the red thick curve.

As can be seen in Figure \ref{fig:rprocess}, the radioactive heating rates exhibit two phases: one in which the value is approximately constant over time (constant phase) and the other in which the value decays with time (decay phase).
It can be seen that the duration of the constant phase has a diversity and tends to be longer for larger values of $\Ye$.
The decay phase can be well described by a power law.
Since the power-law indices are generally smaller than $-1$,
the most amount of radioactive energy is added during the constant phase.
As a first step, we ignore the decay phase (see Section \ref{sec:analytic_bPL} for the case with the decay phase) and 
take the duration and heating rate of the constant phase as parameters and model the heating rate per unit volume (see Equation (\ref{eq:energy})) as follows:
\begin{equation}\label{eq:heating_model}
Q_{\rm heat}=
\left\{
\begin{array}{ll}
\rho \dot{q}_0 & (t<t_{\rm heat})\\
0 & (t>t_{\rm heat})
\end{array}
\right. ,
\end{equation}
where $\dot{q}_0$ is the radioactive energy (except for that in neutrinos) released per unit mass per unit time during the constant phase\footnote{
For the convenience of comparisons with the results by \citet{2010MNRAS.406.2650M} and \citet{2019MNRAS.485.4404D}, we use the unit $\MeVnucs$ ($\sim 10^{18}$ erg g$^{-1}$ s$^{-1}$) for the heating rate $\dot{q}_0$. 
} and $t_{\rm heat}$ is the duration of the constant phase.
The value of $\Ye$ can take a variety of values depending on the outflow mechanism of ejecta as well as the time of ejection.
The values of $\Ye$ in the dynamical ejecta of BNS mergers are about $\Ye\sim0.1$--$0.4$.
As mentioned in Section \ref{sec:initial}, the dominant component of ejecta can be the late-time viscously driven wind.
This component is expected to have a large $\Ye$ compared to that of the dynamical ejecta, with $Y_\mathrm{e} \sim 0.3$--$0.4$, as shown in previous work, e.g., \citet{2020ApJ...901..122F}.
Considering the uncertainties in $\Ye$ over the different components of ejecta, we adopt the simple model described by Equation (\ref{eq:heating_model}) and treat $\dot{q}_0$ and $t_{\rm heat}$ as parameters.
Furthermore, in Section \ref{sec:analytic}, we model the accretion flows semi-analytically.
For this purpose, a simplified treatment in Equation (\ref{eq:heating_model}) is convenient.
The modeling of the case with the heating rates including the decay phase is given in Section \ref{sec:analytic_bPL}.

Here, we neglect a possible effect of heating due to the jet which interacts with the preceding ejecta by making a hot cocoon.
A part of the energy of the jet is converted into the internal energy of the cocoon, so that we can regard the jet as an additional heating source for the ejecta.
However, since this occurs only for a narrow solid angle about the jet axis, the mass of the heated ejecta (i.e., cocoon) is expected to be small relative to the total ejecta mass \citep[e.g.,][]{2021MNRAS.500..627H}.

\subsection{Results}

\begin{figure}[t!]
	\centering
	\includegraphics[width=1.0\linewidth]{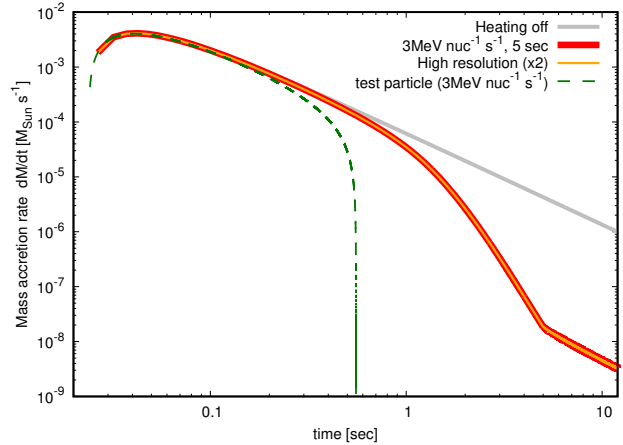}
	\caption{
		Temporal evolution of the mass accretion rate evaluated at $r=650\km$, where $t=0$ corresponds to the beginning of the calculation.
		The thick red curve presents the result for the heating model with $\dot{q}_0 = 3 \MeVnucs$ and $t_{\rm heat}=5\sec$.
		The orange curve is the result for the same model but with the twice finer spatial resolution.
		The relative error between the two different resolutions of calculations is at most 0.2\%.
		The gray curve represents the result for the model without radioactive heating. The green dashed curve represents the mass accretion rate calculated based on the test-particle model (see Appendix \ref{sec:appendix} for details).
	}
	\label{fig:mdot}
\end{figure}

The thick red line in Figure \ref{fig:mdot} presents the time evolution of the mass accretion rate calculated with $\dot{q}_0 = 3\MeVnucs$ and $t_{\rm heat}=5~\sec$ in Equation (\ref{eq:heating_model}).
Here, in order to clearly illustrate the effect of radioactive heating on the mass accretion rate, we have adopted the value of $t_{\rm heat}$ longer than those shown in Figure \ref{fig:rprocess} for the relevant $\dot{q}_0$.
We assume that all the radioactive energies (except for those in neutrinos) are converted to the internal energies of the fluid elements.
The thin gray line shows the fallback rate in the absence of radioactive heating, which can be well described by $dM/dt \propto t^{-5/3}$ (where the initial $\sim 0.05 \sec$ is affected by the initial conditions).
Here, the mass accretion rate is evaluated using the mass flux at $r=650\km$.
It can be seen that the mass accretion rate is substantially suppressed by the effect of radioactive heating.
The break in the red curve at $t=5~\sec$ corresponds to the termination of the heating $t = t_{\rm heat}$.
After $t=t_{\rm heat}$, the mass accretion rate continues roughly in proportion to $t^{-5/3}$.

The dashed green line shows the mass accretion rate calculated using the same method as the test-particle model of \citet{2010MNRAS.406.2650M}.
The detail of the method is described in Appendix \ref{sec:appendix}.
Although the heating rate is the same as that in the fluid model (red line), the time-dependence of the mass accretion rate for the test-particle model differs from that for the fluid model.
In the fluid model, the mass accretion rate does not show a sharp cutoff as observed in the test-particle model at $t \sim 5$ sec but slowly decreases taking a few times longer duration.
This is due to the difference in conversion of the internal energy to the kinetic energy between the test-particle model and the fluid model.
In the test-particle model of \citet{2010MNRAS.406.2650M}, all radioactive energy injected to a fluid element is assumed to be converted into the kinetic energy.
On the other hand, in the fluid model, the radioactive heating first increases the internal energy (or pressure) of a fluid element, and then is converted to the kinetic energy through the pressure gradient.
Actually, in the fluid model, a part of the internal energy from radioactive heating is not converted to the kinetic energy and hence falls to the central object with the fluid element.
As the rate of conversion to kinetic energy is higher, the more amount of ejecta tends to be unbound. 
As a result, the mass accretion rate decreases more slowly in the fluid model than that in the test-particle model.

\begin{figure}[t!]
	\centering
	\includegraphics[width=1.0\linewidth]{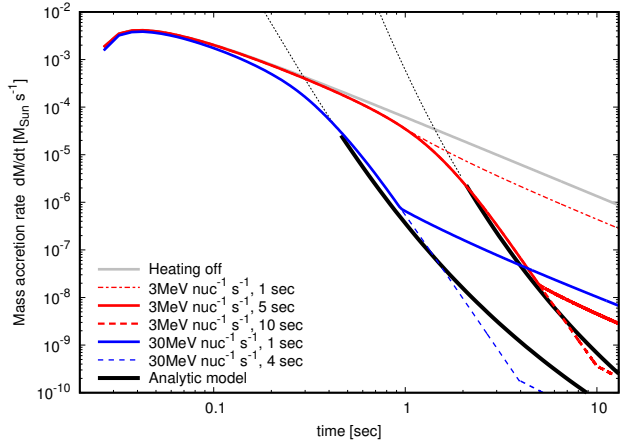}
	\caption{
		Dependence of the mass accretion rate on the heating parameters.
		The gray and red thick curves are the same as those in Figure \ref{fig:mdot}.
		The blue curves represent the case for $\dot{q}_0=30\MeVnucs$.
		The thick, dashed, and dot-dashed curves indicate the results for the different lengths of $t_{\rm heat}$ as shown in the legend.
		The black curves show the mass accretion rates calculated by the semi-analytical model (see Section \ref{sec:analytic} for detail).
		The semi-analytical model is not valid during the times indicated by the black dotted curves.
   }
	\label{fig:mdot2}
\end{figure}

The dependence of the mass accretion rate on the heating parameters is shown in Figure \ref{fig:mdot2}.
The solid red curve is the same as the red curve in Figure \ref{fig:mdot}.
The dot-dashed and thin-dashed curves show the results for $t_{\rm heat}= 1 \sec$ and $t_{\rm heat}= 10 \sec$, respectively.
As can be seen from the figure, the longer the radioactive energy injects, the longer the suppression of the mass accretion rate continues.
The blue curves are those for $\dot{q}_0 = 30\MeVnucs$, where
the solid and dashed curves represent the cases for $t_{\rm heat}=1~\sec$ and $t_{\rm heat}=4~\sec$, respectively.
This value of $\dot{q}_0$ is a factor of a few greater than that reached by radioactive heating (see Figure \ref{fig:rprocess}); we take this value for a possible case with additional energy sources such as shock heating (e.g., due to the subsequent viscosity-driven wind) or strong magnetic field.
It can be seen that the mass accretion rate is suppressed on a shorter timescale than that for $\dot{q}_0 = 3\MeVnucs$ because $\dot{q}_0$ is larger.
The semi-analytical modeling described in Section \ref{sec:analytic} explains these behaviors.

\begin{figure}[t!]
	\centering
	\includegraphics[width=1.0\linewidth]{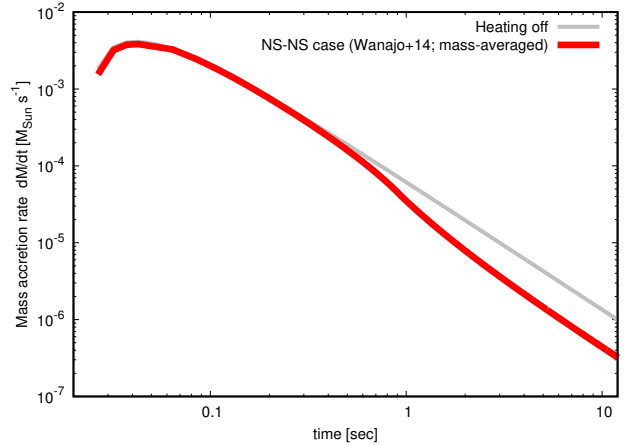}
	\caption{
    Mass accretion rate using the mass-averaged heating rate of the dynamical ejecta calculated by \citet{2014ApJ...789L..39W} (red bold curve).
    The gray curve shows the mass accretion rate without radioactive heating.
    At $10 \sec$, the mass accretion rate with heating is suppressed by about 70\% compared to that without heating.
	}
	\label{fig:mdot_ma}
\end{figure}

Finally, we present the mass accretion rate calculated using a more realistic heating rate rather than using a constant approximation in Equation (\ref{eq:heating_model}).
Figure \ref{fig:mdot_ma} shows the temporal evolution of the mass accretion rate calculated using the mass-averaged heating rate of the dynamical ejecta in \citet{2014ApJ...789L..39W} (see Figure \ref{fig:rprocess}).
At $t=10 \sec$, the mass accretion rate with radioactive heating becomes about 30\% of that without heating.
Although Figure \ref{fig:mdot_ma} only shows the results of numerical calculations until about $10 \sec$, as will be discussed in Section \ref{sec:application}, the mass accretion rate is expected to be suppressed to less than $\sim10$\% of that without heating after $\mathcal{O}(10^4)$--$\mathcal{O}(10^8)$ sec due to the continuous heating in the decay phase.

\section{semi-analytical modeling of fallback accretion}\label{sec:analytic}

\subsection{Characteristic scales of the accretion flow}\label{sec:scale}
For better understanding of the numerical results obtained in Section \ref{sec:CFD}, we develop a semi-analytical modeling.
We first introduce the characteristic scales for the basic Equations (\ref{eq:eoc})--(\ref{eq:eos}).
For this purpose, it is convenient to rewrite Equation (\ref{eq:energy}) into that in terms of $P/\rho$.
By using Equations (\ref{eq:eoc}), (\ref{eq:momentum}), and (\ref{eq:eos}), the energy conservation law (\ref{eq:energy}) can be written as follows:
\begin{equation}\label{eq:nr_eoe}
\frac{\partial}{\partial t}\left(\frac{P}{\rho}\right)
+v \frac{\partial}{\partial r}\left(\frac{P}{\rho}\right)\\
=(\Gamma-1)\left[\dot{q}_0-\frac{P}{\rho} \frac{1}{r^{2}} \frac{\partial}{\partial r}\left(r^{2} v\right)\right].
\end{equation}
The characteristic parameters of the fluid equations (\ref{eq:eoc}), (\ref{eq:momentum}), and (\ref{eq:nr_eoe}) are $GM$ and $\dot{q}_0$.
Hence, the system has the characteristic scales of length and time.
From a dimensional analysis, these can be expressed in the following forms:
\begin{multline}\label{eq:rc}
r_{c}=\left[\frac{(G M)^{3}}{\dot{q}_0^{2}}\right]^{1/5} \\
\sim3540~{\rm km}~\left(\frac{M}{2.7~M_\odot}\right)^{3/5}\left(\frac{\dot{q}_0}{3\MeVnucs}\right)^{-2/5},
\end{multline}
\begin{multline}\label{eq:tc}
t_c=\left[\frac{(G M)^{2}}{\dot{q}_0^{3}}\right]^{1/5}\\
\sim0.35~{\rm s}~\left(\frac{M}{2.7~M_\odot}\right)^{2/5}\left(\frac{\dot{q}_0}{3\MeVnucs}\right)^{-3/5}.
\end{multline}
Combining these quantities, a characteristic scale of velocity is also obtained as
\begin{multline}\label{eq:vc}
v_c=\left(GM\dot{q}_0\right)^{1/5}\\
\sim0.033c~\left(\frac{M}{2.7~M_\odot}\right)^{1/5}\left(\frac{\dot{q}_0}{3\MeVnucs}\right)^{1/5}.
\end{multline}
Then, we introduce the dimensionless length $\xi\equiv r/r_c$, time $\chi\equiv t/t_c$, density $\phi\equiv \rho/\rho_c$, velocity $u\equiv v/v_c$, and pressure $\theta\equiv P/\rho v_c^2=P/\phi\rho_cv_c^2$, where $\rho_c$ is an arbitrary constant with the dimension of density.
Rewriting Equations (\ref{eq:eoc}), (\ref{eq:momentum}) and (\ref{eq:nr_eoe}) by using these dimensionless variables, we obtain the following dimensionless equations:
\begin{equation}\label{eq:nr_eoc_nd}
	\frac{\partial \phi}{\partial \chi}+\frac{1}{\xi^{2}}\frac{\partial}{\partial \xi}\left(\xi^{2} \phi u\right)=0,
\end{equation}
\begin{equation}\label{eq:nr_eom_nd}
	\frac{\partial u}{\partial \chi}+u \frac{\partial u}{\partial \xi}=-\frac{1}{\phi} \frac{\partial \left(\phi\theta\right)}{\partial \xi}-\frac{1}{\xi^{2}},
\end{equation}
\begin{equation}\label{eq:nr_eoe_nd}
		\frac{\partial \theta}{\partial \chi}
	+v \frac{\partial \theta}{\partial \xi}
	=(\Gamma-1)\left[1-\theta\frac{1}{\xi^{2}} \frac{\partial}{\partial \xi}\left(\xi^{2} u\right)\right].
\end{equation}
Because these non-dimensional equations have the same form for any $\rho_c$, there is no characteristic scale of ejecta mass (equivalently mass density) in this system (i.e., Equations (\ref{eq:eoc}), (\ref{eq:momentum}), and (\ref{eq:nr_eoe})).
As mentioned in Section \ref{sec:initial}, we presume that the profile of the viscosity-driven wind can be modeled by enhancing the density $\rho$ (see Equation (\ref{eq:rhomodel})).
The invariance to the density scale ensures that the subsequent semi-analytical model can be used for both the viscosity-driven wind and the dynamical ejecta.

\begin{figure}
	\centering
	\includegraphics[width=1.0\linewidth]{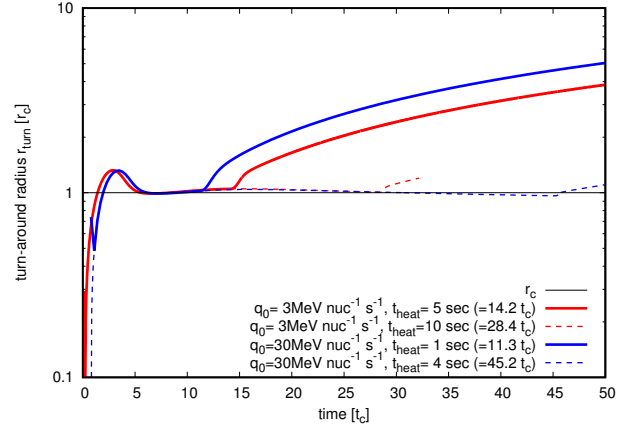}
	\caption{
		Temporal evolution of turn-around radii (where $v=0$).
		The vertical and horizontal axes are normalized by the characteristic scales in Equations (\ref{eq:rc}) and (\ref{eq:tc}), respectively.
		The red and blue curves show the results for $\dot{q}_0=3\MeVnucs$ and $\dot{q}_0=30\MeVnucs$, respectively.
		The thick and dashed curves differ in the length of $t_{\rm heat}/t_c$. 
		The calculation indicated by the red dashed curve is terminated at $t=32\, t_c$.
		Note that the small difference between the red and the blue curves is due to the fact that the initial conditions are not scaled.
	}
	\label{fig:rturn}
\end{figure}

Equations (\ref{eq:eoc}), (\ref{eq:momentum}), and (\ref{eq:nr_eoe}) can be normalized by Equations (\ref{eq:rc})--(\ref{eq:vc}) to eliminate the parameters $GM$ and $\dot{q}_0$.
Therefore, the accretion flow follows the same equations under the characteristic scales (\ref{eq:rc})--(\ref{eq:vc}) up to $t=t_{\rm heat}$.
Figure \ref{fig:rturn} shows the time evolution of the turn-around radius $r_{\rm turn}$, at which the velocity becomes $v=0$.
The vertical axis is normalized by $r_c$ and the horizontal axis by $t_c$.
As can be seen from Figure \ref{fig:rturn}, the difference between these curves is due solely to the difference in $t_{\rm heat}/t_c$.
This clearly shows the effectiveness of the scaling.
It is also seen that from $t=6\, t_c$ to $t_{\rm heat}$, the turn-around radius $r_{\rm turn}$ is approximately constant $r_{\rm turn}\sim r_c$ over time.
Using the chain rule with the differential coefficients of velocity, we can obtain the time evolution of $r_{\rm turn}$ as follows:
\begin{equation}\label{eq:rturn_eq}
	\frac{d r_{\rm turn}(t)}{dt}=-\frac{\left(\partial v/\partial t\right)_{r=r_{\rm turn}}}{\left(\partial v/\partial r\right)_{r=r_{\rm turn}}}.
\end{equation}
At the turn-around radius $r=r_{\rm turn}$, the velocity is $v=0$ by definition, so that the Lagrangian time derivative becomes $Dv/Dt=\partial v/\partial t$.
The fact that $r_{\rm turn}$ is approximately constant over time, $dr_{\rm turn}(t)/dt\sim0$ means $Dv/Dt\sim0$; i.e., at the turn-around radius, the gravity and pressure gradient forces are almost balanced (see Equation (\ref{eq:momentum})).
This indicates that the ejecta outside $r_{\rm turn}$ does not fall back.

\subsection{The hydrodynamical structure inside $r_{\rm turn}$}

\begin{figure}
	\centering
	\includegraphics[width=1.0\linewidth]{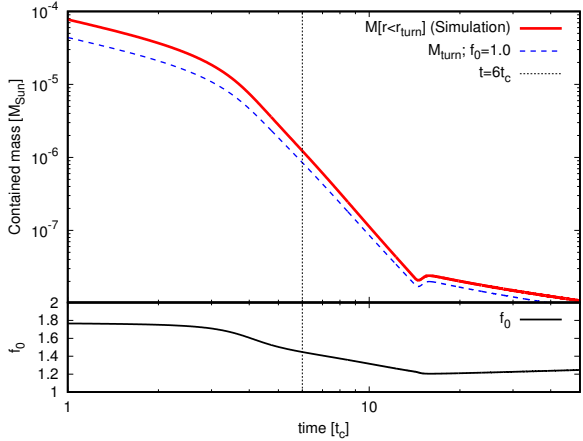}
	\caption{
		The top panel shows the time evolution of the ejecta mass inside the turn-around radius $r_{\rm turn}$ (excluding that within a sufficiently small radius taken to be $r_{\rm count} = 650\km$).
		The horizontal axis is normalized by the characteristic timescale in Equation (\ref{eq:tc}).
		The solid red curve shows the result obtained from the numerical fluid calculation for $\dot{q}_0 = 3\MeVnucs$ and $t_\mathrm{heat} = 5\sec$.
		The blue dashed curve indicates the enclosed mass inside $r_{\rm turn}$ evaluated with $f_0=1$ in Equation (\ref{eq:M_model}), where the values of $r_{\rm turn}$ and $\rho_{\rm turn}$ are adopted from the numerical fluid calculations.
		The bottom panel shows the value of $f_0$ required to reproduce the results of the fluid calculations.
	}
	\label{fig:m_Turn}
\end{figure}

In the following, we discuss the phase after $r_{\rm turn}$ becomes constant, i.e., $t>6~t_c$.
When the position of $r=r_{\rm turn}$ is constant, the forces are balanced (see Equation (\ref{eq:rturn_eq})) and the velocity is $0$; thus the mass does not fall back from the radius beyond $r_{\rm turn}$.
Therefore, the decreasing rate of the ejecta mass between a sufficiently small radius (taken to be $r_{\rm count} = 650\km$ in the numerical calculation) and $r_{\rm turn}$, which has not been accreted yet to the central object, is equal to the mass accretion rate to the central object.
The ejecta mass contained within $r=r_{\rm turn}$ can be written as
\begin{equation}\label{eq:M_model}
{M}_{\rm turn}=\frac{4 \pi}{3}f_{0} r_{\text {turn }}^{3} \rho_{\text {turn }},
\end{equation}
where the subscript ``turn'' means the value at $r=r_{\rm turn}$
and $f_0$ is an $\mathcal{O}(1)$ constant that corrects the difference originating from the mass density distribution.
In Figure \ref{fig:m_Turn}, the enclosed mass within $r=r_{\rm turn}$ evaluated with Equation (\ref{eq:M_model}) is compared to that of the numerical fluid calculation (excluding that within $r_{\rm count}$) for $\dot{q}_0 = 3\MeVnucs$ and $t_\mathrm{heat} = 5\sec$.
The bottom panel of Figure \ref{fig:m_Turn} shows the value of $f_0$ in order for $M_{\rm turn}$ to match the numerical value.
Generally, there are three terms contributing to the time derivative of $M_{\rm turn}$ as follows:
\begin{equation}\label{eq:mmodel_dot}
\frac{\dot{M}_{\rm turn}}{M_{\rm turn}}=
\frac{\dot{f}_0}{f_0}
+3\frac{\dot{r}_{\rm turn}}{r_{\rm turn}}
+\frac{\dot{\rho}_{\rm turn}}{\rho_{\rm turn}}
.
\end{equation}
Evaluating the value of each term from the fluid calculations,
we find that, between $t\sim 6\, t_c$ and $t=t_{\rm heat}$, $r_{\rm turn}$ and $f_0$ vary on longer timescales than the mass density $\rho_{\rm turn}$.
We therefore at first take $r_{\rm turn}$ and $f_0$ to be constants and assume that the time variation of $M_{\rm turn}$ is due only to $\rho_{\rm turn}$.

\begin{figure}
	\centering
	\includegraphics[width=1.0\linewidth]{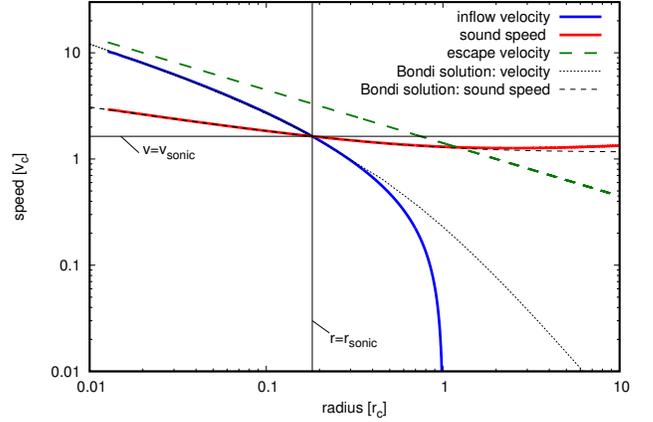}
	\caption{
		The infall velocity $|v|$ (blue line) and the sound speed (red line) of the accretion flow for $\dot{q}_0 = 3\MeVnucs$ and $t_\mathrm{heat} = 5\sec$ at $t=3\sec\sim 8.5\, t_c>6\, t_c$.
		The vertical and horizontal axes are normalized by the characteristic scales in Equation (\ref{eq:vc}) and Equation (\ref{eq:tc}), respectively.
		The black dotted and dashed curves represent the inflow velocity and the sound velocity calculated from the Bondi accretion flow, respectively.
		The green long-dashed curve represents the escape velocity at each radius.
		The vertical and horizontal thin lines mark the sonic radius $r_{\rm sonic}$ and the sonic velocity $v_\mathrm{sonic}$, respectively.
    }
	\label{fig:bondi}
\end{figure}

\begin{figure}
	\centering
	\includegraphics[width=1.0\linewidth]{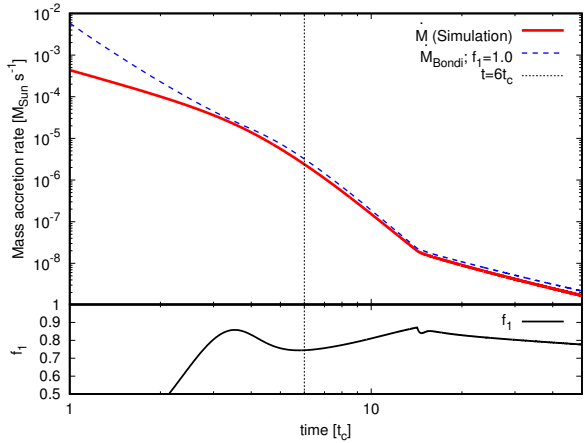}
	\caption{
		The top panel shows the time evolution of the mass accretion rate.
		The horizontal axis is normalized by the characteristic timescale in Equation (\ref{eq:tc}).
		The solid red curve presents the mass accretion rate calculated using the mass flux at $r = 650 \km\sim 0.18 r_c$ in the numerical fluid calculation for $\dot{q}_0 = 3\MeVnucs$ and $t_\mathrm{heat} = 5\sec$.
		The blue dashed curve is evaluated by Equation (\ref{eq:Mdot_model}) with $f_1=1$.
		Here, $\rho_{\rm turn}$ and $a_{\rm turn}$ are the values obtained from the numerical fluid calculation.
		The bottom panel shows the values of $f_1$ required to reproduce the result of the numerical fluid calculation.
	}
	\label{fig:Mdot_bondi}
\end{figure}

The hydrodynamical structure at the time $t=8.5~t_c>6~t_c$, where the turn-around radius $r_{\rm turn}$ has become constant, is shown in Figure \ref{fig:bondi}.
In the region where $|v|\gtrsim |v_c|$, the flow and sound velocities can be approximated well by the Bondi accretion flows.
The accretion rate of the Bondi solution can be written as follows using the physical quantities at $r=r_{\rm turn}$ \citep{1952MNRAS.112..195B}:
\begin{equation}\label{eq:Mdot_model} 
	\dot{M}_{\rm Bondi}=2 \sqrt{2} \pi f_1(G M)^{2} \rho_{\text {turn }} / a_{\text {turn }}^{3},
\end{equation}
where $a=\sqrt{\Gamma P/\rho}$ is the sound speed, $a_{\rm turn}=a(r_{\rm turn})$ and $f_1$ is an $\mathcal{O}(1)$ constant that corrects the difference owing to the fact that $r_{\rm turn}$ is not infinite.
Originally, Equation (\ref{eq:Mdot_model}) is a relation between the gas that is sufficiently distant and stationary, but here we evaluate this value at $r_{\rm turn}$.
The fluid element at the turn-around radius $r_{\rm turn}$ is almost at rest (see Figure \ref{fig:rturn} and Section \ref{sec:scale}).
Although the pressure gradient is comparable to the gravity, the gravitational potential is rather smaller than the internal energy of the fluid, so that it can be approximated in this way.
In order to check the validity of this assumption,
in Figure \ref{fig:Mdot_bondi}, the mass accretion rate obtained by the numerical fluid calculation is compared to that evaluated using Equation (\ref{eq:Mdot_model}).
After $t=4\, t_c$, it can be seen that the mass accretion rate is well approximated by Equation (\ref{eq:Mdot_model}).
Equating the mass accretion rate in Equation (\ref{eq:Mdot_model}) to that in Equation (\ref{eq:mmodel_dot}) (in the absolute values), we obtain the equation for the time evolution of the mass density $\rho$ at $r=r_{\rm turn}$ as follows:
\begin{equation}\label{eq:rho_eq}
	\frac{\dot{\rho}_{\rm turn}}{\rho_{\rm turn}}
	=-\frac{3\sqrt{2}}{2}\frac{f_1}{f_0}\frac{1}{t_c}\left(\frac{a_{\rm turn}}{v_c}\right)^{-3}.
\end{equation}
Here, $f_0$, $f_1$ and $r_{\rm turn}\sim r_c$ are assumed to be constant and Equations (\ref{eq:rc})--(\ref{eq:vc}) are used.
In order to solve Equation (\ref{eq:rho_eq}), a model of the time evolution of the sound speed $a_{\rm turn}$ at $r=r_{\rm turn}$ is needed.

\subsection{Sound speed at $r_{\rm turn}$}\label{sec:sound_speed}

\begin{figure}
	\centering
	\includegraphics[width=1.0\linewidth]{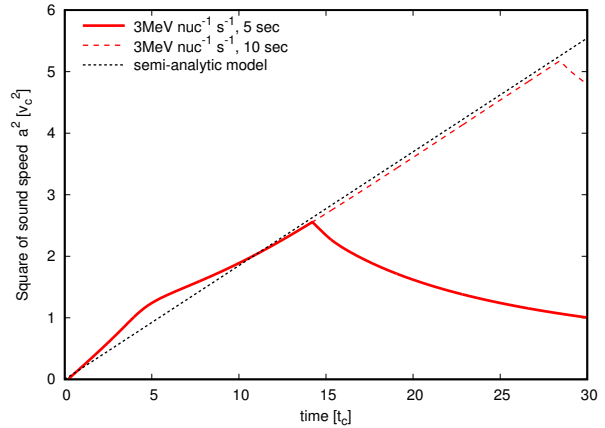}
	\caption{
		Time evolution of the square of the sound speed at $r=r_{\rm turn}$.
		The vertical and horizontal axes are normalized by the characteristic scales in Equation (\ref{eq:vc}) and Equation (\ref{eq:tc}), respectively.
		Note that both the axes are given in linear scales.
		The red solid and dashed curves show the results of the numerical fluid calculations with $\dot{q}_0 = 3\MeVnucs$ for $t_{\rm heat}=5 \sec$ and $t_{\rm heat}=10 \sec$, respectively.
		The black dotted line, which is a linear function of time, is a model curve of the sound speed given by Equation (\ref{eq:sound_speed_model}).
		It can be seen that the model approximates the results of the fluid calculations well at $t \gtrsim 6~t_c$.
	}
	\label{fig:sound}
\end{figure}

In this section, we model the temporal evolution of the sound speed at the turn-around radius $r_{\rm turn}$.
Figure \ref{fig:sound} shows the time evolution of the square of $a_{\rm turn}$.
As can be seen from the figure, the values of $a_{\rm turn}^2$ in the numerical fluid calculations are proportional to time after $t\gtrsim 6\, t_c$ up to $t=t_{\rm heat}$.
This time dependence is expected from the characteristic scale $\dot{q}_0 t$, which has the dimension of the square of the velocity.
We assume the following equation as a model for the temporal evolution of the sound speed at $r=r_{\rm turn}$:
\begin{equation}\label{eq:sound_speed_model}
a_{\rm turn}=\sqrt{\frac{\Gamma P_{\rm turn}}{\rho_{\rm turn}}}
=\beta_0 v_c\left(\frac{t}{t_c}\right)^{1/2},
\end{equation}
where the value of $\beta_0$ includes the effect of adiabatic cooling in the accretion flow (that will be $\sqrt{\Gamma\left(\Gamma-1\right)}\sim 0.67$ in the absence of adiabatic cooling).
Note that, in this paper, we consider the case where the total heating per nucleon is $\mathcal{O}({\rm MeV})$, so that this value will never reach the speed of light.
By solving Equation (\ref{eq:sound_speed_model}) for $P_{\rm turn}$ and by using Equations (\ref{eq:eos}), (\ref{eq:tc}), and (\ref{eq:vc}),
we obtain the value of the internal energy, $\epsilon_{\rm int}\sim 0.42\, \rho\dot{q}_0 t$, which indicates that about half of the added radioactive energy is converted to internal energy and the rest to kinetic energy.
The test-particle model of \citet{2010MNRAS.406.2650M} and \citet{2019MNRAS.485.4404D} assumes that all of the radioactive energy is converted into kinetic energy; however as we have shown here, \wiadd{such a 100\% conversion efficiency from internal energy to kinetic energy is not the case.}

\subsection{Halting of the fall-back accretion by $r$-process}\label{sec:rhalt_result}

The mass accretion rate $\dot{M}_{\rm turn}$ at $r=r_{\rm turn}$, can be calculated if the time evolution of $\rho_{\rm turn}$ is obtained with Equations (\ref{eq:rho_eq}) and (\ref{eq:sound_speed_model}).
Using Equations (\ref{eq:M_model}), (\ref{eq:rho_eq}), and (\ref{eq:sound_speed_model}), we obtain the following relation:
\begin{equation}\label{eq:mdot_index}
	\frac{d \ln \dot{M}_{\rm turn}}{d\ln t}
	=-\frac{3}{2}\left[1+\sqrt{2}\frac{f_1}{f_0}\frac{1}{\beta_0^3}\left(\frac{t}{t_c}\right)^{-1/2}\right].
\end{equation}
By integrating this over time, we obtain
\begin{equation}\label{eq:mdot_analytic}
	\frac{\dot{M}_{\rm turn}}{\dot{M}_0}=\left(\frac{t}{t_0}\right)^{-3/2}\exp\left[-\frac{3\sqrt{2}}{\beta_0^3}\frac{f_1}{f_0}\sqrt{\frac{t_c}{t_0}}\left(1-\sqrt{\frac{t_0}{t}}\right)\right],
\end{equation}
where $\dot{M}_0$ is the initial condition for $\dot{M}_{\rm turn}$ at a given time $t_0$.
For the evaluation of $f_0$, $f_1$, and $\beta_0$, we adopt the values at $t_0=6\, t_c$, at which the turn-around radius becomes approximately constant (see Figure \ref{fig:rturn}), and use $f_1/f_0=0.5$ and $\beta_0=0.43$.
The resultant mass accretion rates using Equation (\ref{eq:mdot_analytic}) are shown in Figure \ref{fig:mdot2} by black curves.
The dotted curves indicate those for $t<6\, t_c$, i.e., the range of time when the assumptions necessary to derive equation (\ref{eq:mdot_analytic}) are not valid.
It is clear from Figure \ref{fig:mdot2} that the mass accretion rate can be well approximated by Equation (\ref{eq:mdot_analytic}) after $t=6\, t_c$, which decreases rapidly along the theoretical curve.
Although the model is calibrated based on the result for $\dot{q}_0=3\MeVnucs$, it can be seen that the theoretical curve also explains well those for $\dot{q}_0=30\MeVnucs$.

 $\dot{M}_0$ is a free parameter that cannot be determined due to the lack of a typical scale of ejecta mass in this system.
For the cases in which only $\dot{q}_0$ is different but the ejecta profile is the same, we can derive a scaling law of $\dot{M}_0$ for various $\dot{q}_0$.
For $t\ll6\, t_c$, the mass accretion rate exhibits almost the same temporal evolution as in $\dot{q}_0=0$ case (see the gray line in Figure \ref{fig:mdot2}), so that we can approximate the mass accretion rate as that without radioactive heating.
Because the ejecta mass follows the same dimensionless equation under the normalized scales given by Equations (\ref{eq:rc})--(\ref{eq:vc}),
the ratio of mass accretion rates between with and without heating, which is a dimensionless quantity, has the same time evolution as a function of the normalized time $t/t_c$ for various $\dot{q}_0$.
If we choose the reference point $t_0$ in time units of $t_c$ (as we chose the reference point of $\dot{M}_0$ as $t_0=6\, t_c$), this ratio at $t=t_0$ is uniquely determined.
Without heating, the mass accretion rate has the time evolution proportional to $t^{-5/3}$ so that the scaling law for $\dot{M}_0$ with respect to $\dot{q}_0$ is given as follows:
\begin{equation}\label{eq:mdot0_dep}
	\dot{M}_0\propto t_c^{-5/3}\propto \left(GM\right)^{-2/3}\dot{q}_0.
\end{equation}
In the setting of this paper, at $t=6\, t_c$, the value of $\dot{M}_0$ is about 13\% of that without heating, which is derived from the numerical result with $\dot{q}_0=3\MeVnucs$. 
In the case of $\dot{q}_0=30 \MeVnucs$ in Figure \ref{fig:mdot2}, we adopt $\dot{M}_0$ calculated using Equation (\ref{eq:mdot0_dep}). 

As represented by Equation (\ref{eq:mdot_analytic}) and Figure \ref{fig:mdot2}, the mass accretion rate is suppressed compared to the case without radioactive heating.
This can be understood from the following two effects by considering the mass accretion rate $\dot{M}\propto a_{\rm B}r_{\rm B}^2\rho_{\rm B}$ evaluated at the sonic point $r=r_{\rm B}$.
The first effect is that $r_{\rm turn}$ is nearly constant over a certain period of time.
As mentioned in Section \ref{sec:scale}, this corresponds to the fact that no mass can flow into the inside of the sphere of $r_{\rm turn}$ from the outside.
Thus the mass density $\rho_{\rm turn}$ at the turn-around radius will necessarily decrease with accretion.
Furthermore, the flow outside the sonic radius $r_{\rm B}$ is almost incompressible because it is a subsonic flow, so that the mass density $\rho_{\rm B}$ at the sonic point is comparable to $\rho_{\rm turn}$.
Therefore, $\rho_{\rm B}$ is also a decreasing function of time.
The other effect is the increase in the speed of sound with time.
As can be seen in Figure \ref{fig:bondi}, the speed of sound at the sonic point of the accretion flow is about the same order of magnitude as that at the turn-around radius.
Since the speed of sound increases in proportion to $t^{1/2}$ (see Equation (\ref{eq:sound_speed_model})), the radius of the sonic point $r_{\rm B}=GM/2a_{\rm B}^2$ \citep{1952MNRAS.112..195B} shrinks as $r_{\rm B}\propto t^{-1}$.
Because the mass accretion rate is $\dot{M}\propto a_{\rm B}r_{\rm B}^2\rho_{\rm B}\propto \rho_{\rm B} t^{-3/2}$, the above two effects both work to reduce the accretion rate.

\citet{1989ApJ...346..847C} calculated the accretion rate with radioactive heating, mainly due to $^{56}$Ni, in the context of fallback accretion to the proto-neutron star in a supernova explosion.
He showed analytically that the fallback accretion rate declines sharply as $\dot{M}\propto t^{-9/2}$ well within the half-life of $^{56}$Ni, i.e., for a period of time when the heating rate per unit mass is approximately constant.
Although the solution we obtained is exponential rather than the power-law of time (see Equation (\ref{eq:mdot_analytic})), the result in \citet{1989ApJ...346..847C} is qualitatively similar to ours in terms of the suppression of the mass accretion rate in the presence of heating.
The difference is that the model in \citet{1989ApJ...346..847C} considered only a self-similar expansion of ejecta and assumed the mass density decreasing as $t^{-3}$.
In fact, provided that $\rho_B\propto t^{-3}$, we obtain $\dot{M}\propto\rho_{\rm B} t^{-3/2}\propto t^{-9/2}$ by using the same argument as described earlier.
The difference arises because we also consider the reduction of mass density $\rho_{\rm B}$ due to accretion.

\begin{figure*}[!t]
	\centering
	\includegraphics[width=1.0\linewidth]{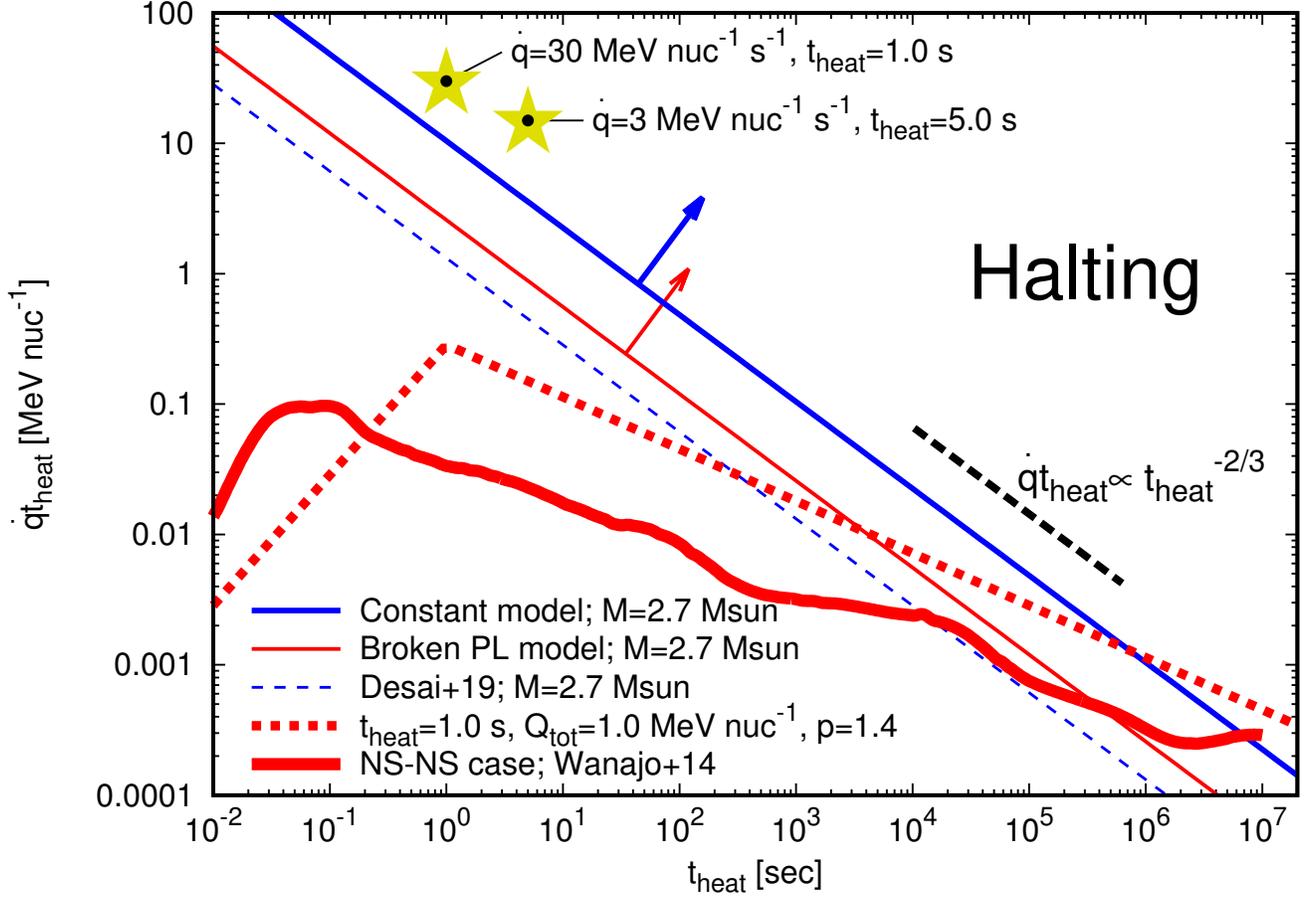}
	\caption{
        The parameter regions for radioactive heating, in which the mass accretion is sufficiently suppressed.
        The solid blue and red lines represent the boundaries of the suppression for the case of a constant heating rate (see Equation (\ref{eq:halting_condition})) and the case of a broken-power-law heating rate as expressed in Equation (\ref{eq:heating_PLC}) ($q_0=3.0 \MeVnucs$, $t_{\rm heat}=1.0~\sec$, and $p=1.4$), respectively, for a central mass of $2.7 M_{\odot}$.
        \wiadd{
        The thick black dash lines are shown to guide the dependence of the halting condition $\dot{q}\propto t_{\rm heat}^{-5/3}$ (see Equations (\ref{eq:halting_condition}) and (\ref{eq:heating_PLC})).
        }
		In the upper-right regions above these lines, the mass accretion is well suppressed by the effect of radioactive heating.
		The star-shaped symbols indicate the locations for the parameters used in the numerical calculations of this paper, for which the halting has been observed.
        The thin blue dashed line shows the boundary for the ``cutoff condition'' by \citet{2019MNRAS.485.4404D} (see Appendix for details).
		The thick red line is the locus of the mass-averaged heating rate (excluding that in neutrinos) calculated by \citet{2014ApJ...789L..39W} (see Figure \ref{fig:rprocess}).
		The thick red dotted line is that for the model in which the total radioactive energy per nucleon is $1\MeVnuc$, the constant phase lasts for $1\sec$, and the heating rate decreases proportionally to $t^{-1.4}$ in the decay phase.
	}
	\label{fig:cutoff_condition}
\end{figure*}

If $t_{\rm heat}$ is shorter than $6~t_c$, the stagnation of $r_{\rm turn}$ and the increase of the sound speed will not occur for the constant heating model.
Thus, $t\gtrsim 6~t_c$ is the necessary condition for these two processes to work, namely,
\begin{multline}\label{eq:halting_condition}
	t_{\rm heat} > K t_c\\
	\sim 2.1 \sec~K_{\rm 6}\left(\frac{M}{2.7~M_\odot}\right)^{2/5}\left(\frac{\dot{q}_0}{3\MeVnucs}\right)^{-3/5},
\end{multline}
where $K$ is the time (with respect to $t_c$) that takes for $r_{\rm turn}$ to become constant and $K=6\, K_6$ (see Figure \ref{fig:rturn}).
After $t\sim 6~t_c$, the analytic solution (\ref{eq:mdot_analytic}) becomes valid, and the mass accretion rate rapidly decreases.
Therefore, we call this time, $t_{\rm halt}\sim 6~t_c$, ``halting time'' and this suppression of mass accretion ``halting''.
The halting time corresponds to the timescale in which the mass accretion rate decreases to $\sim13$\% of its original value (see the description below Equation (\ref{eq:mdot0_dep})).
For given $\dot{q}_0$, $t_{\rm heat}$, and $M$, the halting condition is determined by Equation (\ref{eq:halting_condition}).
Figure \ref{fig:cutoff_condition} shows whether or not the halting occurs in the $\dot{q}_0t_{\rm heat}$--$t_{\rm heat}$ space.
Here, because we are considering only the constant heating phase (see Equation (\ref{eq:heating_model}), and see also Equation (\ref{eq:heating_PLC}) for the case with a power-law decay phase), the vertical axis in Figure \ref{fig:cutoff_condition} indicates the total radioactive energy injected into the fluid.
As can be seen from Figure \ref{fig:cutoff_condition}, even if the total radioactive energy is the same,
the halting is more likely to occur as $t_{\rm heat}$ increases.
This is because the ejecta which is accreted at the later phase has less binding energy, and therefore, the accretion is more easily disturbed by the later injection of radioactive energy.

\citet{2019MNRAS.485.4404D} also argued that the mass accretion stops at a finite time if there is sufficient heating.
We compare our model with that of \citet{2019MNRAS.485.4404D} as shown by the thin dashed lines in Figure \ref{fig:cutoff_condition}.
The method for calculating their theoretical curve is summarized in Appendix \ref{sec:appendix}.
We consider that the ``halting time'' we obtained corresponds to the ``cutoff'' of the mass accretion claimed by \citet{2019MNRAS.485.4404D}.
As can be seen from the figure, their results and ours have the same dependence on the variables, but \wiadd{the total heating energy per nucleon $\dot qt_\mathrm{heat}$ to halt the fallback accretion} differ by about an order of magnitude for the constant heating case.
In other words, compared to their test-particle model, our fluid model requires about $10$ times larger heating rate to halt the mass accretion for given $t_{\rm heat}$.
This difference is likely due to the fact that in the test-particle model, all of the radioactive energy is converted into the kinetic energy, whereas in the fluid model, this energy conversion is incomplete remaining internal energy.

\section{Semi-analytical model with power-law decaying heating rate}\label{sec:analytic_bPL}

As we find in Figure \ref{fig:rprocess}, there is actually not only a constant phase in the heating profile but also a decay phase.
The heating profile can be reasonably approximated by
\begin{equation}\label{eq:heating_PLC}
	\dot{q}(t)=\left\{
	\begin{array}{ll}
		\dot{q_0} & (t\leq t_{\rm heat,0})\\
		\dot{q_0}\left(\frac{t}{t_{\rm heat,0}}\right)^{-p} & (t>t_{\rm heat,0})
	\end{array}
	\right.,
\end{equation}
where $t_{\rm heat,0}$ is the duration of the constant phase and $p > 1$.
The total radioactive energy $Q_{\rm tot}$ can be written as
\begin{equation}\label{eq:BPL_total}
	Q_{\rm tot}=\frac{p}{p-1}\dot{q_0}t_{\rm heat,0}.
\end{equation}
If the halting occurs within the constant phase of radioactive heating in Equations (\ref{eq:heating_PLC}), that is $t_{\rm halt}\leq t_{\rm heat,0}$, the halting time $t_{\rm halt}=Kt_c$ (cf. Equation (\ref{eq:halting_condition})) can be written as follows:
\begin{multline}\label{eq:halting_time_const}
	t_{\mathrm{halt}}
	= \left[K^{5}(G M)^{2} \left(\frac{p}{p-1}\right)^3 Q_{\rm tot}^{-3}t_{\mathrm{heat},0}^{3}\right]^{\frac{1}{5}} \\
	\sim 265 \sec K_6\left(\frac{M}{2.7 M_{\odot}}\right)^{2/5}
	\left(\frac{Q_{\rm tot}}{1 \MeVnuc}\right)^{-3/5}\\
	\times \left(\frac{t_{\mathrm{heat}, 0}}{300\sec}\right)^{3/5}
	\ \ (p=1.4).
\end{multline}

Even if the mass accretion does not halt during the constant phase, the halting will eventually occur.
Here, we construct a semi-analytical model of the accretion flow with the heating that decays with the power-law of time.
In the decay phase, the typical scales given by Equations (\ref{eq:rc}) and (\ref{eq:tc}) become time-varying as follows:
\begin{equation}\label{eq:rc_var}
\begin{aligned}
r_{c}(t)=\left[\frac{(G M)^{3}}{\dot{q}(t)^{2}}\right]^{1 / 5}=\left[\frac{(G M)^{3}}{\dot{q}_{0}^{2}}\right]^{1 / 5}\left(\frac{t}{t_{\text {heat,0}}}\right)^{2 p / 5},
\end{aligned}
\end{equation}
\begin{equation}\label{eq:tc_var}
t_{c}(t)=\left[\frac{(G M)^{2}}{\dot{q}(t)^{3}}\right]^{1 / 5}=\left[\frac{(G M)^{2}}{\dot{q}_{0}^{3}}\right]^{1 / 5}\left(\frac{t}{t_{\text {heat,0}}}\right)^{3 p / 5}.
\end{equation}
Introducing the dimension-less variables $\xi=r/r_c(t)$ and $\chi=t/t_c(t)$, the fluid equations (\ref{eq:eoc})--(\ref{eq:energy}) can be normalized as follows:
\begin{multline}\label{eq:eoc_bPL}
\left(1-\frac{3 p}{5}\right) \frac{\partial \ln \phi}{\partial \ln \chi}
+\left(V-\frac{2 p}{5}\right) \frac{\partial \ln \phi}{\partial \ln \xi}\\
+\frac{\partial V}{\partial \ln \xi}
+3 V=0,
\end{multline}
\begin{multline}\label{eq:eom_bPL}
\left(1-\frac{3 p}{5}\right) \frac{\partial V}{\partial \ln \chi}
+\left(V-\frac{2 p}{5}\right) \frac{\partial V}{\partial \ln \xi}\\
+\frac{\partial Z}{\partial \ln \xi}
+Z \frac{\partial \ln \phi}{\partial \ln \xi}
+2 Z+V(V-1)
=-\frac{\chi^{2}}{\xi^{3}},
\end{multline}
\begin{multline}\label{eq:eoe_bPL}
\left(V-\frac{2 p}{5}\right) \frac{\partial Z}{\partial \ln \xi}
+\left(1-\frac{3 p}{5}\right) \frac{\partial Z}{\partial \ln \chi}\\
+(\Gamma-1) Z \frac{\partial V}{\partial \ln \xi}
+[(3 \Gamma-1) V-2] Z
=(\Gamma-1) \frac{\chi^{3}}{\xi^{2}},
\end{multline}
where $\phi=\rho/\rho_c$ is the normalized mass density, $\rho_c$ is arbitrary constant with dimension of mass density, $V=vt/r$ is the normalized radial velocity, and $Z=(P/\rho) (t/r)^2$ is the normalized pressure.
Note that the normalized equations (\ref{eq:eoc_bPL})--(\ref{eq:eoe_bPL}) do not explicitly include $t_{\rm heat,0}$.
As we will see below, $t_{\rm heat,0}$ is relevant to the evolution of ejecta only as an initial condition in Equations (\ref{eq:eoc_bPL})--(\ref{eq:eoe_bPL}).

The temporal evolution of the accretion flow under the heating rate (\ref{eq:heating_PLC}) is as follows.
Up to time $t_{\rm heat,0}$, as seen in Section \ref{sec:analytic}, the accretion flow evolves according to Equations (\ref{eq:nr_eoc_nd})--(\ref{eq:nr_eoe_nd}) with the normalization of Equations (\ref{eq:rc}) and (\ref{eq:tc}).
If $t_{\rm heat,0}$ is longer than $\sim6~t_c$ (see Section \ref{sec:rhalt_result}), the halting occurs during the constant phase, and the halting time is expressed by Equation (\ref{eq:halting_time_const}).
On the other hand, if $t_{\rm heat,0}\lesssim 6~t_c$, the ejecta evolves according to Equations (\ref{eq:nr_eoc_nd})--(\ref{eq:nr_eoe_nd}) and (\ref{eq:eoc_bPL})--(\ref{eq:eoe_bPL}), respectively, before and after
\begin{equation}\label{eq:chi_heat}
\chi_{\rm heat}=t_{\rm heat,0}/t_c(t_{\rm heat,0}),
\end{equation}
at which the heating rate switches from the constant phase to the power-law decaying phase.
As seen in Section \ref{sec:analytic}, during the constant phase, since the temporal evolution of the ejecta with various parameters are identical under normalized variables, almost independently of the initial conditions.
Therefore, even if model parameters are different but $\chi_{\rm heat}$ and $p$ are the same, these accretion flows follows the same temporal evolution according to Equations (\ref{eq:eoc_bPL})--(\ref{eq:eoe_bPL}) from the same initial conditions (i.e., the states at the end of the constant phase) with normalized variables.

In the case of the halting in the decay phase, unlike the case in the constant phase, the normalized halting time $K$ takes a different value from $6$ (see Section \ref{sec:analytic}), where $K$ can be written as 
\begin{equation}\label{eq:chi_halt}
K=t_{\rm halt}/t_c(t_{\rm halt}).
\end{equation}
As seen in Section \ref{sec:rhalt_result}, the halting time $t_{\rm halt}$ is the time it takes for the mass accretion rate with the heating to be suppressed to about $13\%$ of that without heating.
As mentioned earlier, the normalized halting time $K$ is basically a function of $\chi_{\rm heat}$ and $p$ only.
Also, if $t_{\rm heat,0}$ (or $\chi_{\rm heat}$) is long enough (i.e., $\chi_{\rm heat}>6$), this will match the constant model, and $K=6$.
We investigate the dependence of $K$ on $\chi_{\rm heat}$ and $p$ based on the numerical calculation.
As a heating rate profile, we take various values of $t_{\rm heat,0}$ and $p$ in Equation (\ref{eq:heating_PLC}).
We choose $\dot{q}_0$ and $t_{\rm heat,0}$ for $p=1.2$, $1.3$, and $1.4$, respectively, 
with the condition $\dot{q}_0=2.0 \left(t_{\rm heat,0}/1.0~\sec\right)^{-p} \MeVnucs$ for various $t_{\rm heat,0}$.
The ejecta profiles (see Figure \ref{fig:ejecta}) and the calculation method are the same as in Section \ref{sec:CFD}.

\begin{figure}[!t]
	\centering
	\includegraphics[width=1.0\linewidth]{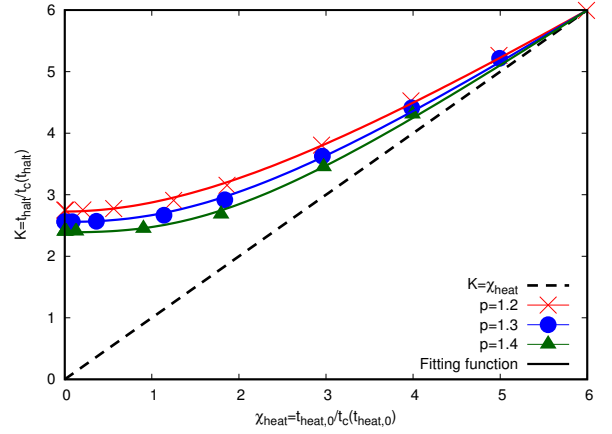}
	\caption{
	Dependence of the normalized halting time $K$ on the normalized duration $\chi_{\rm heat}$ of the constant phase and the decay index $p$ of the heating profile.
	Here we adopt the mass of the central object $M=2.7~M_\odot$, and the parameters of the heating rate with the condition $\dot{q}_0=2.0 \left(t_{\rm heat,0}/1.0~\sec\right)^{-p} \MeVnucs$ for various $t_{\rm heat,0}$.
    The red, blue, and green points show the results of numerical calculations with $p=1.2$, $1.3$, and $1.4$, respectively.
    The solid curves show the fitting functions in Equations (\ref{eq:K_fitting}) and (\ref{eq:K_fitting_paras}) for various $p$.
    The black dash line represents the line of $K=\chi_{\rm heat}$.
    For $\chi_{\rm heat}=6$, the condition $K=6$ shown in Section \ref{sec:analytic} is recovered because the mass accretion halts within the constant phase.
    }
    \label{fig:KvsKheat}
\end{figure}

Figure \ref{fig:KvsKheat} shows the dependence of the normalized halting time $K$ on the normalized duration $\chi_{\rm heat}$ of the constant phase.
The red, blue, and green dots represent the results of the numerical calculation for $p=1.2$, $1.3$, and $1.4$, respectively.
We fit these results with the following monotonically increasing functional form:
\begin{equation}\label{eq:K_fitting}
K^{\alpha}=K_{0}^{\alpha}+\left[1-\left(\frac{K_{0}}{6}\right)^{\alpha}\right] \chi_{\rm heat}^{\alpha},
\end{equation}
where
\begin{equation}\label{eq:K_fitting_paras}
K_0=A_K p+B_K,~~~\alpha=C_K p+D_K.
\end{equation}
Here we fix the value of $K=6$ for $\chi_{\rm heat}=6$ to recover the result of the model with the constant heating exactly.
The resultant parameters are $A_K=-1.68$, $B_K= 4.75$, $C_K=-3.14$, and $D_K=-1.80$.
The resultant fitting functions are shown as solid curves in Figure \ref{fig:KvsKheat}.
Using the radioactive heating rate in the BNS merger as shown in Figure \ref{fig:rprocess}, we find that for that heating rate, $K=2.6$ (with $\chi_{\rm heat}=0$ and $p=1.3$) is appropriate for almost all $\Ye$ cases.
Once we obtain the value of $K$, we can calculate the halting time in the decay phase by solving the equation,
\begin{equation}\label{eq:halting_time_def}
t_{\rm halt}=K\left[\frac{\left(GM\right)^2}{\dot{q}(t_{\rm halt})^3}\right]^{1/5},
\end{equation}
and the halting time is obtained as follows,
\begin{multline}\label{eq:halting_time}
	t_{\mathrm{halt}} =\left[K^{5}(G M)^{2} \left(\frac{p}{p-1}\right)^3 Q_{\rm tot}^{-3} t_{\mathrm{heat},0}^{3(1-p)}\right]^{\frac{1}{5-3 p}}\\ \sim 3.9\times 10^{3} \sec K_{2.6}^{6.25}\left(\frac{M}{2.7 M_{\odot}}\right)^{2.5}
	\left(\frac{Q_{\rm tot}}{1 \MeVnuc}\right)^{-3.75}\\
	\times \left(\frac{t_{\mathrm{heat}, 0}}{1\sec}\right)^{-1.5}
	 \ \ (p=1.4).
\end{multline}
Here we adopt $K = 2.6$ corresponding to $\chi_{\rm heat}\sim 0$, which is a good approximation for the case of realistic heating rates shown in Figure \ref{fig:rprocess}.
This formula holds if the halting does not occur during the constant phase, and thus the above expression is only valid if $t_{\rm halt}$ is longer than $t_{\rm heat,0}$.
Note that, as can be seen from Figure \ref{fig:cutoff_condition}, the power-law index of time in the decay phase of the heating rate $\dot{q}(t)$ must be shallower than $t^{-5/3}$ for halting to occur.

Let us evaluate the mass accretion rate in the decay phase of the heating $\dot{q}(t)\propto t^{-p}$ at $t>t_{\rm halt}$.
Here, we assume that $p<5/3$ so that the halting occurs during the decay phase.
From Equation (\ref{eq:rc_var}), the turn-around radius $r_{\rm turn}$ is expected to evolve in time as $r_{\rm turn}\propto \dot{q}(t)^{-2/5}\propto t^{2p/5}$.
Furthermore, from Equation (\ref{eq:sound_speed_model}), the sound speed at turn-around radius is expected to be $a_{\rm turn}\propto (\dot{q}(t)t)^{1/2}\propto t^{(1-p)/2}$.
We have checked that these time dependencies are approximately consistent with those obtained by numerical calculations, except for the constant factor (e.g., $r_{\rm turn}/r_{\rm c}(t)\sim 0.8$ rather than unity).
By solving Equation (\ref{eq:M_model}) for $\rho_{\rm turn}$ and substituting it into Equation (\ref{eq:Mdot_model}), we obtain the differential equation for the mass $M_{\rm turn}$ within the turn-around radius,
\begin{equation}
\frac{d \ln M_{\rm turn}}{dt}=\frac{3\sqrt{2}}{2}f_{\rm PL}\frac{f_1}{f_0}\frac{\left(GM\right)^2}{a_{\rm turn}^3r_{\rm turn}^3},
\end{equation}
where $r_{\rm turn}$ depends on time.
We find that $f_0$, $f_1$, $\beta_0$, and $r_{\rm turn}/r_c$ in the power-law heating model slightly deviate from those in the constant heating model shown in Section \ref{sec:analytic}, and depend on the normalized duration of the constant phase $\chi_{\rm heat}$ and the decay index of the heating rate $p$.
Here, we introduce the time-constant factor $f_{\rm PL}=f_{\rm PL}(\chi_{\rm heat}, p)$ to adjust these differences.
By integrating this equation over time and differentiating the obtained solution with $t$, we get
\begin{multline}\label{eq:mdot_analytic_decay}
\frac{\dot{M}_{\rm turn}}{\dot{M}_0}=\left(\frac{t}{t_{\rm halt}}\right)^{(3p-15)/10} \\
\times\exp\left[-\frac{5}{5-3p}f_{\rm PL}\frac{3\sqrt{2}}{\beta_0^3}\frac{f_1}{f_0}\frac{1}{\sqrt{K}}\left(1-\left(\frac{t_{\rm halt}}{t}\right)^{\frac{5-3p}{10}}\right)\right].
\end{multline}
It can be easily checked that Equation (\ref{eq:mdot_analytic}) is recovered when $p=0$ and $f_{\rm PL}=1$.
From the calculation results used for determining the $\chi_{\rm heat}$ dependence of $K$, we can also obtain the fitting function of $f_{\rm PL}$.
We fit the calculation results of $f_{\rm PL}$ with the following functional form:
\begin{equation}\label{eq:fpl_fitting}
f_{\rm PL}=\left(S_{\rm PL}p+T_{\rm PL}\right)\chi_{\rm heat}^2+\left(U_{\rm PL}p+V_{\rm PL}\right).
\end{equation}
The resultant parameters are $S_{\rm PL}=-1.22\times 10^{-2}$, $T_{\rm PL}= 2.65\times 10^{-2}$, $U_{\rm PL}=-0.324$, and $V_{\rm PL}=0.661$.

\begin{figure}[!t]
	\centering
	\includegraphics[width=1.0\linewidth]{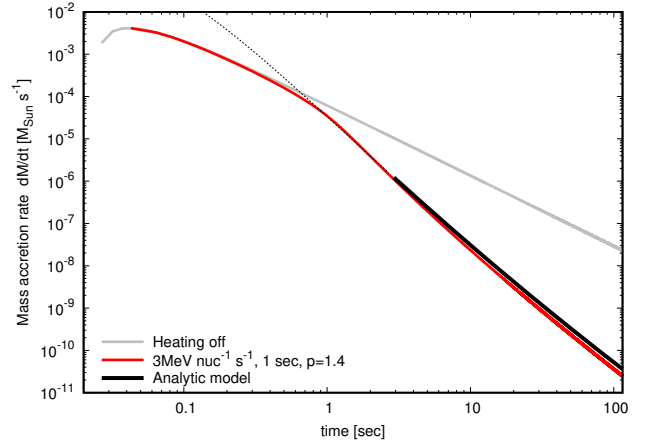}
	\caption{
	Mass accretion rate calculated using the heating rate profile of broken power-law (Equation (\ref{eq:heating_PLC})) with $q_0=3\MeVnucs$, $t_{\rm heat}=1.0~\sec$, and $p=1.4$ (red curve).
    The gray curve shows the mass accretion rate without heating.
    The black curve is the mass accretion rate calculated by the semi-analytical model of Equation (\ref{eq:mdot_analytic_decay}).
    The dotted line is before the halting time, and the Equation (\ref{eq:mdot_analytic_decay}) is not valid.
    }
	\label{fig:Mdot_bPL}
\end{figure}

In Figure \ref{fig:Mdot_bPL}, the numerical calculation result with the parameters $q_0=3\MeVnucs$, $t_{\rm heat,0}=1.0~\sec$, and $p=1.4$ (shown in the red curve) is compared with the mass accretion rate calculated from Equation (\ref{eq:mdot_analytic_decay}) (shown in the black curve).
Note that the numerical results shown in Figure \ref{fig:Mdot_bPL} are for parameters independent of the calculations used to calibrate Equations (\ref{eq:K_fitting_paras}) and (\ref{eq:fpl_fitting}).
As can be seen from the figure, the semi-analytical model reproduces the numerical results well.
Furthermore, generally, with the parameters used in our calculations (i.e., $f_{\rm PL}\sim 0.3$, $f_1/f_0\sim 0.5$ and $\beta_0\sim0.43$), 
the mass accretion rate in Equation (\ref{eq:mdot_analytic_decay}) decreases more sharply than $t^{-5/3}$ at $t=t_{\rm halt}$, so that the mass accretion is expected to be strongly suppressed after the halting time even in the decay phase.

\section{Application to BNS mergers}\label{sec:application}

In this section, we consider the halting process with the realistic heating rate in BNS mergers as shown in Figure \ref{fig:rprocess}.
We can estimate the halting time from the intersection of the heating rate curve and the line above which halting occurs (red solid line in Figure \ref{fig:cutoff_condition} for $M=2.7 M_{\odot}$ and $K\sim 2.6$).
As shown by the thick red curve in Figure \ref{fig:cutoff_condition}, when using the mass-averaged heating rate of the nucleosynthesis calculations in Figure \ref{fig:rprocess}, we can see that the halting occurs by the heating only after a time $\sim 3.7\times 10^5$ sec.

\begin{figure*}
	\centering
	\includegraphics[width=1.0\linewidth]{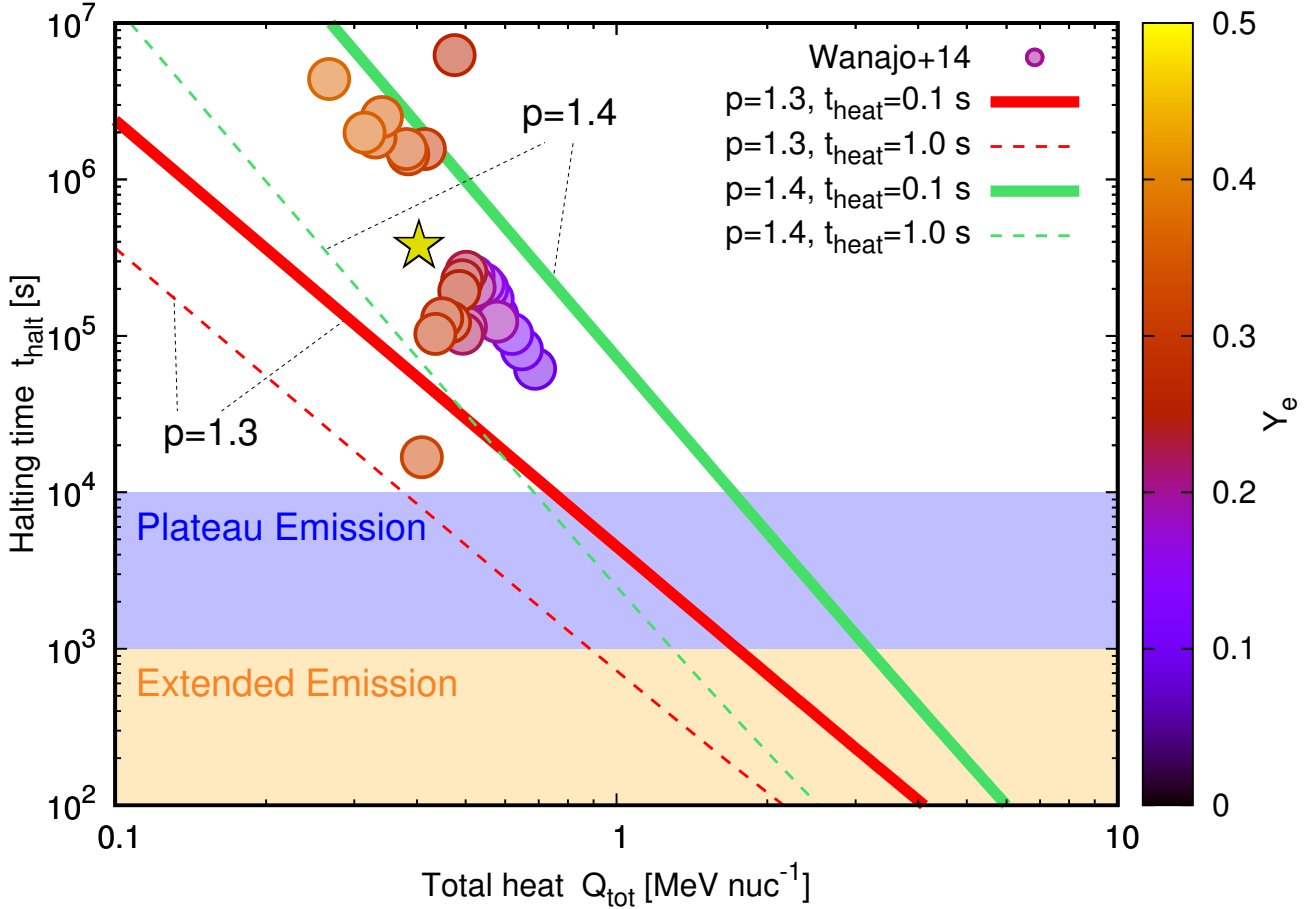}
	\caption{
		Dependence of the halting time on the radioactive heating for the fallback accretion with the central object mass $M=2.7 M_{\odot}$.
		The circles indicate the halting times calculated by using the nucleosynthesis results for individual initial $Y_\mathrm{e}$ values in \citet{2014ApJ...789L..39W}.
		The star represents the results for the mass-averaged heating rate (see Figure \ref{fig:rprocess}).
		The red and green lines represent the results of the calculations with $p=1.3$ and $p=1.4$ power-law index in the decay phase of radioactive heating (see Equation (\ref{eq:heating_PLC})), respectively, for $t_{\rm heat} = 0.1$ sec (solid) and $t_{\rm heat} = 1.0$ sec (dotted).
		Note that, in all cases, the halting occurs in the decay phase.
		Shaded areas are the ranges of typical observed timescales for the extended emission and the plateau emission of sGRBs.
		We can see that \wiadd{the accretion ceases, long after these emissions decay.}
	}
	\label{fig:halting_time}
\end{figure*}

Figure \ref{fig:halting_time} shows the dependence of the halting time on the parameters of the heating, i.e., $t_{\rm heat}$ and $Q_{\rm tot}$.
In addition, in order to investigate the dependence on $\Ye$, we also show the halting times calculated using the heating rates in Figure \ref{fig:rprocess}.
Reflecting the tendency of larger $Q_{\rm tot}$ for smaller $\Ye$, the halting time becomes shorter for smaller $\Ye\lesssim0.3$.
For larger $\Ye$, the dependence of the halting time on $\Ye$ is not monotonic.
This is because as $\Ye$ increases, the length of $t_{\rm heat}$ increases.
The longer $t_{\rm heat}$ leads to the greater heating rate in the decay phase, so that the halting time becomes shorter.
We find that the halting by $r$-process does not occur within a timescale of $\sim 10^4$ sec, although the radioactive heating may decelerate the accretion flow to some extent.

The radioactive heating of ejecta is the thermalization process of non-thermal particles due to ionization losses of charged particles and repeated scattering and absorption of gamma-rays.
It has been pointed out that, after about some of ten days, the timescale of thermalization becomes longer than the expansion timescale of ejecta and thus the thermalization becomes inefficient \citep{2016MNRAS.459...35H,2016ApJ...829..110B,2018MNRAS.481.3423W,2019ApJ...876..128K,2020ApJ...891..152H}.
Because the halting time we obtained is 10 days or even longer, we expect that the thermalization is actually insufficient.
However, all the charged particles associated with radioactive decay, which are efficiently trapped inside the ejecta by the magnetic field, contribute to pressure being independent of their thermalization.
Since the halting is essentially due to an increase in pressure rather than an increase in temperature, the thermalization efficiency is less important for charged particles.
On the other hand, gamma-rays (about a half of $\beta$-decay energy is emitted in the form of gamma-rays) can escape from the ejecta.
Thus, gamma-rays make little contribution to pressure after a sufficient time and the heating rate becomes effectively small.
As can be seen from Equation (\ref{eq:halting_time}), when the heating rate becomes about twice as small, the halting time becomes about an order of magnitude longer.
Therefore, the halting time shown in Figure \ref{fig:halting_time} should be considered as the lower limit.
Note that, if the thermalization is insufficient, the radiation efficiency is small, so that the radiative cooling is negligible.

The uncertainty in nuclear physics may also affect the estimated value of the halting time.
According to \citet{2020arXiv201011182B}, there is a systematic variation of about one order of magnitude in the radioactive heating rate at $\mathcal{O}(1)$--$\mathcal{O}(10)$ days depending on the adopted nuclear ingredients.
Considering this uncertainty in the heating rate and the inefficient thermalization, the halting time for a mass-averaged heating rate is around $\mathcal{O}(10^4)$--$\mathcal{O}(10^8) \sec$.
Note that, according to \citet{2021ApJ...906...94Z}, the uncertainty in the heating rate after $\mathcal{O}(10)~{\rm days}$ become larger, and they suggested the uncertainty range is about three orders of magnitude at $\mathcal{O}(10^8)~\sec$, allowing for a possibility of the halting time larger than $\mathcal{O}(10^8)~{\rm sec}$.

Typical observed timescales for the extended emission and the plateau emission are about $\mathcal{O}(10^3) \sec$ and $\mathcal{O}(10^4) \sec$, respectively \citep{2013MNRAS.430.1061R,2019ApJ...877..147K}.
As can be seen in Figure \ref{fig:halting_time}, the estimated halting time is much longer than these timescales, when using the nucleosynthesis results (\citet{2014ApJ...789L..39W}; see Figure \ref{fig:rprocess}).
If the energy source for these emissions were purely the fallback accretion in BNS mergers \citep{2007MNRAS.376L..48R,2009MNRAS.392.1451R,2017ApJ...846..142K},
the radioactive heating from decaying $r$-process nuclei appears insufficient to disturb the energy supply from the accretion flow.
Some additional heating sources which inhibit accretion or other mechanisms, for example, the time-varying radiation efficiency of the accreting matter \citep[e.g.,][]{2013ApJ...766...31K},
may be required to explain the characteristic timescales of the extended and plateau emissions.

\section{Conclusions}\label{sec:summary}
The discovery of GRB 170817A associated with GW170817 established that a BNS merger is a source of sGRBs.
However, the origin of the late-time emission in sGRBs, namely, extended emission and plateau emission, is still unknown.
We have investigated the fallback accretion model of these long-lasting emission of sGRBs \citep[e.g.,][]{2007MNRAS.376L..48R,2007NJPh....9...17L,2009MNRAS.392.1451R,2015ApJ...804L..16K,2017ApJ...846..142K}.
While the canonical fallback accretion rate of $t^{-5/3}$ has no typical timescales, \citet{2010MNRAS.406.2650M} and \citet{2019MNRAS.485.4404D} discussed that the effect of radioactive heating results in a timescale of $\mathcal{O}(10)$--$\mathcal{O}(100) \sec$ for mass accretion by using a test-particle model.
We have revisited the effect of the radioactive heating due to decaying $r$-process nuclei on the fallback accretion by using a hydrodynamic model rather than a test-particle model.
We have shown that the timescale for the suppression of mass accretion becomes an order of magnitude longer than that in the test-particle model.
Furthermore, we have found no temporal gap (i.e., halt and revival) of mass accretion, being opposed to the results by \citet{2010MNRAS.406.2650M} and \citet{2019MNRAS.485.4404D}.
Their model assumes that all of the radioactive energy is promptly converted to the kinetic energy.
In addition, they also assume that a fluid element does not fall back once it becomes unbound.
However, our fluid calculations have revealed that these assumptions are inappropriate.

We have developed a semi-analytical model for the temporal evolution of mass accretion (see Equation (\ref{eq:mdot_analytic}) and (\ref{eq:mdot_analytic_decay})), which reproduces the numerical results.
The fallback accretion has characteristic length and time scales that depend on the mass of the central object and the radioactive heating rate (see Equations (\ref{eq:rc}) and (\ref{eq:tc})).
Normalizing the hydrodynamical equations with these scales, we have obtained the scale-free equations (see Equations (\ref{eq:nr_eoc_nd})--(\ref{eq:nr_eoe_nd})).
Semi-analytical modeling of these normalized equations has allowed us to investigate a wide parameter range of accretion flow.
While the radioactive heating with a constant rate continues, the radius at which the accretion flow stagnates (turn-around radius) becomes nearly a constant value, being approximately equal to the characteristic length scale.
We have found that the accretion flow inside the turn-around radius can be well approximated by the Bondi accretion flow, and that the mass accretion rate is well reproduced by the Bondi accretion rate evaluated at the turn-around radius.
Furthermore, we have derived the conditions on the heating rate for the substantial suppression of mass accretion (see Equation (\ref{eq:halting_condition})).
We have extended this condition to more general heating profiles that decay with time (see Figure \ref{fig:cutoff_condition}).
For the case where the heating rate can be written as a combination of constant and decay phases (see Equation (\ref{eq:heating_PLC})), we have found that as long as the heating rate decays more slowly than $t^{-5/3}$ in the decay phase, the halting will occur after a sufficient amount of time (see Equation (\ref{eq:halting_time})).
For typical BNS mergers (GRB 170817-like events), the halting timescale for the suppression of mass accretion is found to be $\mathcal{O}(10^4)$--$\mathcal{O}(10^8)$ sec, which is, however, much longer than the timescales in the late-time activity of sGRBs $\mathcal{O}(10^2)$--$\mathcal{O}(10^4)$ sec \citep{2013MNRAS.430.1061R,2017ApJ...846..142K,2019ApJ...877..147K}.
The observations of macronovae/kilonovae associated with sGRBs suggest that the amount and distribution of $r$-process product differ from event to event \citep{2018ApJ...860...62G,2019MNRAS.486..672A}.
For events such as GW170817, where the macronova/kilonova light curve can be observed in detail, the halting time will be determined by modeling the heating rate in the same way.
Even if the detailed light curves can not be obtained, by estimating the abundance distribution (such as the lanthanide fraction) from color evolution, we may be able to obtain the halting time from $\Ye$ as seen in Figure \ref{fig:halting_time}.
Besides, our model will be applicable not only to BNS mergers but also to the fallback accretion of proto-neutron stars in supernova explosions with r-process nucleosynthesis \citep[e.g.,][]{2006ApJ...642..410N,2015ApJ...810..109N,2015Natur.528..376M}.

Our results imply the existence of different mechanisms or different sources of heating, which can stop the late-time activity of sGRBs.
For example, the shock heating by the interaction between the viscously driven wind and the accretion flow may occur.
In order to examine such a mechanism, multi-dimensional hydrodynamical simulations with the effects of radioactive heating will be necessary, in which both outflow (such as late-time viscously driven wind) and inflow (such as the fallback accretion of early dynamical ejecta) exist \citep[see][for a recent development]{2020arXiv201214711K}.
Note that, for a system with only inflow, as in our calculation, the multidimensionality has a minor effect.
In the time evolution of the mass accretion rate, the ejecta profile near the boundary between the gravitationally bound and unbound states is essential.
As seen in Figure \ref{fig:ejecta}, the radial dependence of velocity and density around the radius $r\sim490\km$ (the boundary of the bound ejecta) is independent of latitude, which justifies the calculation with the spherically symmetric model.
Alternatively, magnetic reconnection or other magnetic field dissipation processes may play a role in heating in the ejecta.
Instead of invoking other heating sources, the time scales of extended and plateau emission may be explained by considering a mechanism in which the conversion rate from gravitational energy to radiation decreases rapidly.
For instance, there may be a rapid change in radiation efficiency due to the state transition of the accretion disk as the accretion rate decreases over time \citep[e.g.,][]{2013ApJ...766...31K}.
We leave these issues for our future work.

The halting time is sensitive to the uncertainty of the radioactive heating rate in the $r$-process elements, which ranges from $10^4$--$10^8$ sec for one order of magnitude ambiguity in the heating rate \citep{2020arXiv201011182B}.
Furthermore, it has been suggested that the uncertainty becomes larger in the later stages ($\mathcal{O}(1)$--$\mathcal{O}(10)\yr$) \citep{2021ApJ...906...94Z}.
This indicates that, if we can obtain the halting time for a macronova/kilonova event, we may be able to constrain the physical conditions for the $r$-process as well as the relevant nuclear ingredients.
One possible observational sign is the X-ray excess in the yearly-scale light curve of GW170817 \citep[e.g.,][]{2019ApJ...886L..17H,2021arXiv210304821B,2021arXiv210402070H}, which we are currently investigating \citep{2021ApJ...916L..13I}.

\section*{Acknowledgments}
We thank the anonymous referee for fruitful comments.
We are grateful to 
Masaru Shibata, 
Kazuya Takahashi, Hamidani Hamid, Tomoki Wada, Koutarou Kyutoku,
Sho Fujibayashi, Kyohei Kawaguchi, Hiroki Nagakura, 
Shota Kisaka, Kazumi Kashiyama, and Shuta Tanaka
for fruitful discussion and valuable comments.
We thank the participants and the organizers of the workshops with the identification number YITP-T-19-04, YKIS2019 and YITP-T-20-19 for their generous support and helpful comments.
This work is supported by Grants-in-Aid for Scientific
Research No. 21J01450 (WI), 18H01213 (KK), 20H01901, 20H01904, 20H00158, 18H01213, 18H01215, 17H06357, 17H06362, 17H06131 (KI) from the Ministry
of Education, Culture, Sports, Science and Technology (MEXT) of Japan.

\appendix


\section{Test-particle model}\label{sec:appendix}
\subsection{Cold case}
First, we describe the fallback accretion in a system with negligible pressure.
The velocity of a fluid particle evolves over time according to the equation of motion:
\begin{equation}\label{EoM}
\frac{dv}{dt}=-\frac{GM}{r^2},
\end{equation}
where $M$ is the mass of the central object.
The first integral of Equation (\ref{EoM}) gives the dynamical energy per mass of the fluid element, which is written as
\begin{equation}
E_0=\frac{1}{2}v^2-\frac{GM}{r}.
\end{equation}
Let us introduce dimensionless variables $x=r/r_s$ and $\beta=v/c$, where $r_s=2GM/c^2$.
The dimensionless time is also defined as $\tau=t/t_s$, where $t_s=r_s/c$.
The dimensionless energy per mass $\lambda$ is determined by
\begin{equation}
\lambda\equiv\frac{1}{x_0}-\beta_0^2=\frac{1}{x}-\beta^2=-2E_{0}/c^2,
\end{equation}
where the subscript $0$ indicates the values in the initial state.

The turn-around time, i.e., the time it takes for the fluid particle to change the direction of motion, can be written as:
\begin{equation}\label{eq:turnaround}
\tau_f=\frac{\beta_0}{\lambda\left(\beta_0^2+\lambda\right)}+\tan^{-1}\left(\frac{\beta_0}{\sqrt{\lambda}}\right)\lambda^{-3/2}.
\end{equation}
Note that $\tau_f$ is a function of the initial velocity and energy.
When $\lambda\ll\beta_0^2<1$, $\tau_f$ can be written as:
\begin{equation}\label{tauf_late}
\tau_f\sim\frac{\pi}{2}\lambda^{-3/2}.
\end{equation}
As can be seen from Equation (\ref{tauf_late}), the turn-around time of a marginally bound fluid particle ($\lambda\sim 0$) hardly depends on the initial velocity.

Let us find the mass per unit time, which falls back through the sphere of $r=r_{\rm fin}$.
Since the dynamical energy $E_0$ is conserved, the time it takes to return to $r_{\rm fin}$ from the point at which $v=0$ coincides with $\tau_f\left(r_{\rm fin},E_0(r_0)\right)$.
Therefore, the time $t_{\rm fb}$ required for the fluid particle launched at a velocity $v_0$ from a radius $r_0$ to fall back to $r_{\rm fin}$ can be written as follows:
\begin{equation}
t_{\rm fb}(r_0,E_0(r_0))=t_s\left[\tau_f(r_0,E_0(r_0))+\tau_f\left(r_{\rm fin},E_0(r_0)\right)\right].
\end{equation}
Once the fallback time is determined as a function of $r_0$,
the mass accretion rate $\dot{M}$ is calculated as follows:
\begin{equation}\label{Mdot_calc}
\dot{M}\left(t_{\mathrm{fb}}\left(r_{0}, E_{0}\left(r_{0}\right)\right)\right)\equiv
\frac{d M}{d t_{\mathrm{fb}}}
=4 \pi r_{0}^{2} \rho_{0}\left(r_{0}\right)\left(\frac{d t_{\mathrm{fb}}}{d r_{0}}\right)_{r=r_{0}}^{-1},
\end{equation}
where $\rho(r_0)$ is the mass density in the initial state.

\subsection{Test-particle model for the $r$-process halting}

According to \citet{2010MNRAS.406.2650M} and \citet{2019MNRAS.485.4404D}, we calculate the fallback time with radioactive heating per unit time, $\dot{q}$.
Assuming that all the radioactive energy is converted to the kinetic energy,
the total energy of the particle at the turn-around time can be estimated as follows:
\begin{equation}\label{eq:Ef_eq}
E_f(r_0)=E_0(r_0)+\int_{t_{\text {start}}}^{t_s\tau_{f}\left(r_0, E_f\right)}\dot{q}(t)\, dt,
\end{equation}
where the subscript $f$ represents the values in the final state, i.e., the values at the turn-around radius.
Here, in order to introduce the effect that the turn-around time becomes longer as the energy of the particle increases,
the value of $\tau_f$ at the upper end of the integration is evaluated by using $E_f$.
In fact, the internal energy injected to the fluid element is converted into the kinetic energy via the pressure gradient forces.
Since it is difficult to deal with this process in the test-particle model, we adopt Equation (\ref{eq:Ef_eq}), which is the same prescription in \citet{2010MNRAS.406.2650M} and \citet{2019MNRAS.485.4404D}.
Furthermore, according to \citet{2019MNRAS.485.4404D}, after the turn-around time (or, equivalently, fluid particles with $v<0$), we neglect the effect of radioactive heating.
Therefore, the fallback time is written as:
\begin{equation}\label{tfb_wheat}
t_{\text {fb}}\left(r_{0}\right)=
t_s\left[
\tau_f\left(r_{0}, E_{f}\left(r_{0}\right)\right)+
\tau_f\left(r_{\text {fin}}, E_{f}\left(r_{0}\right)\right)
\right].
\end{equation}
Using Equations (\ref{Mdot_calc}) and (\ref{tfb_wheat}), the mass accretion rate $\dot{M}$ when radioactive heating is effective can be obtained.

\subsection{Halting condition for the test-particle model}
\begin{figure}
	\centering
	\includegraphics[width=1.0\linewidth]{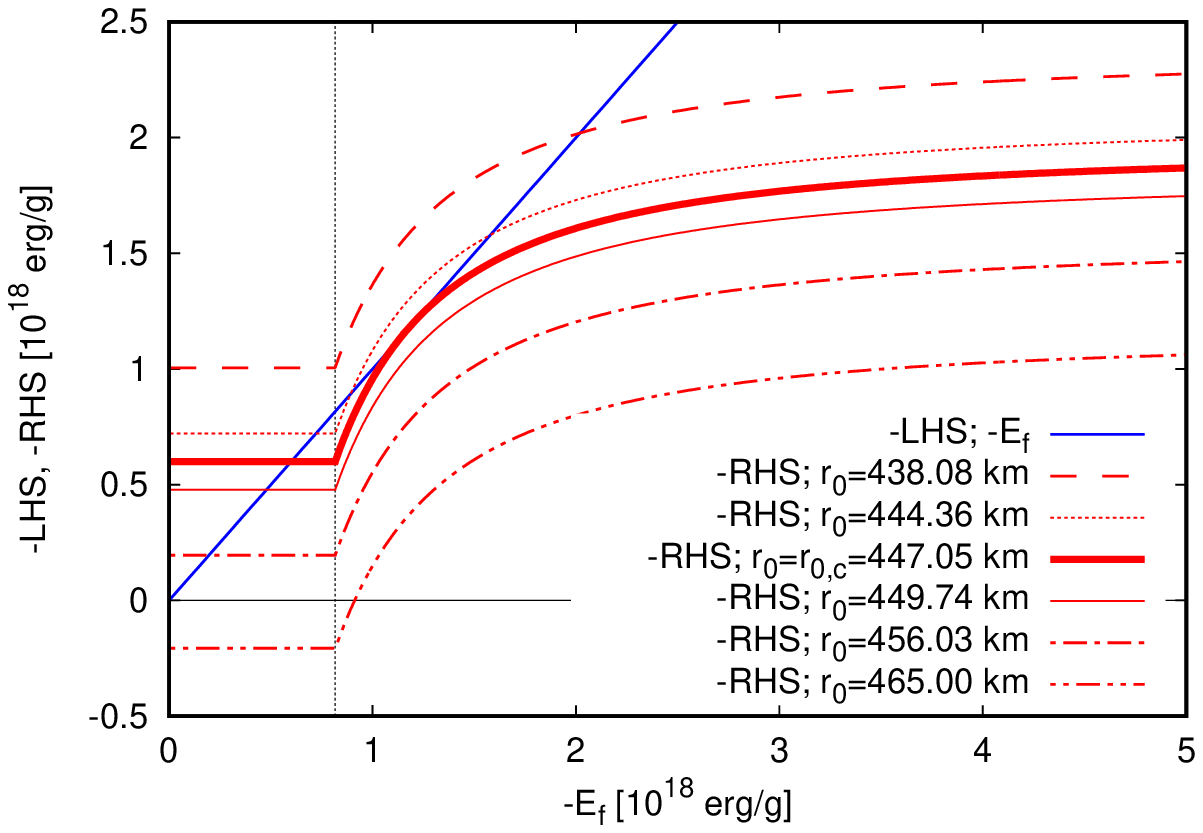}
	\caption{
    Left (blue line) and right (red curves) hand sides (with negative signs) of the algebraic Equation (\ref{eq:const_tp}) as a function of $-E_f$ in the test-particle model.
    The parameters $\dot{q}_0=3\MeVnucs$ and $t_{\rm heat}=0.6\sec$ are adopted, which correspond to a case such that mass accretion stops and resumes, i.e., making a ``gap'' \citep[for details see text below Equation (\ref{eq:t_resume}) or][]{2010MNRAS.406.2650M}.
    The vertical dotted line indicates the value of $E_f$ such that the turn-around time becomes equal to $t_{\rm heat}$.
    The different types of red curves represent differences in the initial positions of the fluid particles, with the upper lines corresponding to the inner initial positions, i.e., closer to the central object.
    The thick red line indicates the case of the initial radius $r_{0,c}$ being the boundary at which the solution series vary discontinuously.
  }
	\label{fig:tp_model}
\end{figure}

Let us analytically evaluate the test-particle model for the case in which $\dot{q}$ is constant with time.
Considering only marginally bound ejecta, we evaluate Equation (\ref{tfb_wheat}) by using Equation (\ref{tauf_late}).
In this case, Equation (\ref{eq:Ef_eq}) can be written as an algebraic equation for $E_f$ as follows:
\begin{equation}\label{eq:const_tp}
	E_f(r_0)=
	\left\{
	\begin{array}{ll}
		\displaystyle E_0(r_0)+\dot{q}_0t_s\frac{\pi}{2}\left(-\frac{2E_f}{c^2}\right)^{-3/2} & (t_s\tau_f<t_{\rm heat}) \vspace{1mm}\\
		\displaystyle E_0(r_0)+\dot{q}_0t_{\rm heat} & (t_s\tau_f\geq t_{\rm heat})
	\end{array}
	\right.
	,
\end{equation}
where we assume $t_{\rm start}\ll t_s\tau_f$.
This can be calculated by finding the intersection of the two curves represented by the right-hand side (red) and the left-hand side (blue) as shown in Figure \ref{fig:tp_model}.
If there are multiple intersections, the solution with the smallest $E_f$ (i.e., the solution with the shortest turn-around time) is the physical solution.

The series of solutions discontinuously change at which Equation (\ref{eq:const_tp}) has a double root.
Since $E_{f,c}$, which corresponds to the point such that the blue line comes into contact with the red thick curve in Figure \ref{fig:tp_model}, is the root of the derivative of Equation (\ref{eq:const_tp}), it can be written as follows:
\begin{equation}
	E_{f,c}=-\frac{1}{2}\left({3\pi\dot{q}_0GM}\right)^{2/5}.
\end{equation}
The corresponding fallback time can be written as
\begin{equation}
	t_{\rm fb,c}\sim2t_s\tau_f\left(r_{0,c},E_{f,c}\right)
	\sim \left(\frac{32}{27}\pi^2\right)^{1/5}t_c,
\end{equation}
or numerically,
\begin{equation}\label{eq:tp_cutoff_time_num}
	t_{\rm fb,c}\sim 1.64t_c
	\sim0.58\sec \left(\frac{Q_{\rm tot}}{3\MeVnuc}\right)^{-3/5} \left(\frac{M}{2.7~M_{\odot}}\right)^{2/5} \left(\frac{t_{\rm heat}}{1\sec}\right)^{3/5}.
\end{equation}
Here we used $Q_{\rm tot}\sim\dot{q}_0t_{\rm heat}$, assuming that the offset $t_{\rm start}$ of the calculation start time is sufficiently shorter than $t_{\rm heat}$.
The corresponding radius $r_{0,c}$ is determined from
\begin{equation}
	E_0(r_{0,c})=-\frac{5}{6}\left(3\pi\dot{q}_0GM\right)^{2/5}.
\end{equation}
The right-hand side of this equation corresponding to the solution ($r_{0,c}, E_{f,c}$) is shown as the thick red curve in Figure \ref{fig:tp_model}.
As can be seen from the figure, the velocities of fluid particles released from the radius smaller than $r_{0,c}$ become $v=0$ before reaching $t= t_{\rm heat}$ and then the particles start infalling.
On the other hand, the fluid particles released from the radius greater than $r_{0,c}$, which corresponds to those represented by the red curves below the thick red curve, continue to be heated until $t= t_{\rm heat}$.
In order for a fluid particle released from the radius greater than $r_{0,c}$ to have a bound solution (i.e., $E_f<0$), the following condition is required:
\begin{equation}
E_0(r_{0,c})+\dot{q}_0t_{\rm heat}<0.
\end{equation}
Rewriting this condition in terms of $t_{\rm heat}$ and $Q_{\rm tot}$, we obtain
\begin{equation}\label{eq:gap_UL}
	t_{\rm heat} < \sqrt{\frac{3125}{3456}}\pi t_{s}\left(\frac{Q_{\mathrm{tot}}}{c^{2}}\right)^{-3 / 2}.
\end{equation}
If this condition is satisfied, even a particle released from the outside of the sphere of $r_{0,c}$ by an infinitesimal distance (see the red curve for $r_0=449.74\km$) has a finite fallback time $t_{\rm fb,r}$ longer than $t_{\rm fb,c}$, namely,
\begin{equation}\label{eq:t_resume}
t_{\rm fb,r}=\frac{3}{4}\left(\frac{\pi^4}{3}\right)^{1/10}\left(\frac{5}{4}-\frac{t_{\rm heat}}{t_{\rm fb,c}}\right)^{-3/2}t_c.
\end{equation}
Evaluating the value of $t_{\rm fb,r}$ for $\dot{M}=2.7M_\odot$, $\dot{q}_0=3\MeVnucs$, and $t_{\rm heat}=0.6 \sec$ gives $t_{\rm fb,r}\sim3.95 \sec$.
Further, outwardly released fluid particles (see the red curve of Figure \ref{fig:tp_model} for $r_0=456.03\km$) have a longer fallback time than $t_{\rm fb,r}$ and thus the mass accretion continues.
This is exactly the ``gap'', the suspension of mass accretion between $t= t_{\rm fb,c}$ and $t_{\rm fb,r}$, which has been shown in \citet{2010MNRAS.406.2650M}.
On the other hand, if $t_{\rm heat}$ is sufficiently long such that Equation (\ref{eq:gap_UL}) is not satisfied, the mass accretion halts and never resumes.
This is what has been demonstrated as a ``cutoff'' case in \citet{2010MNRAS.406.2650M}.
In fact, we find a cutoff at the time calculated from Equation (\ref{eq:tp_cutoff_time_num}) for the test-particle model shown in Figure \ref{fig:mdot}.

Equation (\ref{eq:gap_UL}) is only a necessary condition for which a gap of mass accretion appears.
For this case, there must be a double root $r_{0,c}$, in other words, $t_{\rm heat}$ must be sufficiently long enough for mass accretion to stop once.
This can be given by $t_s\tau_f\left(r_{0,c},E_{f,c}\right)>t_{\rm heat}$.
This can be also written as a condition for $t_{\rm heat}$ and $Q_{\rm tot}$:
\begin{equation}\label{eq:gap_LL}
	\frac{1}{6 \sqrt{3}} \pi t_{s}\left(\frac{Q_{\text {tot }}}{c^{2}}\right)^{-3 / 2}<t_{\text {heat }}.
\end{equation}

A gap of mass accretion appears in the test-particle model if both Equations (\ref{eq:gap_UL}) and (\ref{eq:gap_LL}) are satisfied.
\citet{2010MNRAS.406.2650M} classified the qualitative behavior of mass accretion by introducing a parameter $\eta\equiv t_{\rm heat}/t_{\rm fb,c}$.
Using this parameter, we have
\begin{equation}
	\frac{1}{2}<\eta<\frac{5}{4}.
\end{equation}
The lower and upper limits of the inequality represent the conditions under which mass accretion stops and resumes, respectively.
The upper bound of $1.25$ was also obtained in \citet{2019MNRAS.485.4404D}, which confirms our theoretical framework being equivalent to theirs.
The cutoff condition represented by the dashed lines for the test-particle model in Figure \ref{fig:cutoff_condition} is determined such that the lower bound of $\eta$ becomes $1/2$.


\bibliography{sample631}{}
\bibliographystyle{aasjournal}



\end{document}